\documentclass[%
 aip,
 jcp,%
 amsmath,amssymb,
reprint,%
]{revtex4-1}

\usepackage{graphicx}
\usepackage{dcolumn}
\usepackage{bm}

\begin{document}

\author{Jakub K. Sowa}
\affiliation
{Department of Materials, University of Oxford, OX1 3PH Oxford, UK.}
\altaffiliation{Current address: Department of Chemistry, Rice University, Houston, TX 77005, USA.} 
\email{jakub.sowa@rice.edu}

\author{Rudolph A. Marcus}
\affiliation
{Noyes Laboratory of Chemical Physics, California Institute of Technology, Pasadena, CA 91125, USA}
\email{ram@caltech.edu}

\title[ ]
  {On the theory of charge transport and entropic effects in solvated molecular junctions}

\begin{abstract}
Experimental studies on single-molecule junctions are typically in need of a simple theoretical approach that can reproduce or be fitted to experimentally measured transport data. In this context, the single-level variant of the Landauer approach is most commonly used but methods based on Marcus theory are also gaining in popularity.
Recently, a generalized theory unifying these two approaches has also been developed. 
In the present work, we extend this theory so that it includes entropic effects (which can be important when polar solvents are involved, but are likely minor for solid-state systems). We investigate the temperature-dependence of the electric current and compare it to the behavior predicted by the Landauer and the conventional Marcus theory. 
We argue that this generalized theory provides a simple yet effective framework for understanding charge transport through molecular junctions.
Furthermore, we explore the role of the entropic effects in different transport regimes and suggest experimental criteria for detecting them in solvated molecular junctions.
Lastly, in order to account for nuclear tunnelling effects, we also demonstrate how lifetime broadening can be introduced into the Marcus-Levich-Dogonadze-Jortner-type description of electron transport.
\end{abstract}

\maketitle
\section{Introduction \label{Intro}}
Following a series of experimental breakthroughs that took place around the turn of the millennium,\cite{reed1997conductance,park2000nanomechanical,park2002coulomb,liang2002kondo,cui2001reproducible,kergueris1999electron} the field of molecular electronics has seen two decades of rapid experimental and theoretical development. From the technological perspective, the focus has been largely put on proof-of-principle experiments. It has been shown, for instance, that electronic devices based on molecular junctions can act as transistors,\cite{park2000nanomechanical,perrin2015single,gehring2017distinguishing} rectifiers,\cite{elbing2005single,diez2009rectification,perrin2016gate} spintronic devices\cite{iacovita2008visualizing,sanvito2011molecular,bogani2008molecular} or thermoelectric materials.\cite{reddy2007thermoelectricity,cui2017perspective} These experimental studies were performed in a multitude of device geometries and on a plethora of molecular structures.
Currently, however, progress beyond such prototypical devices is also slowly being made. It has been demonstrated, for example, that it is possible  to construct molecular diode devices based on self-assembled molecular monolayers which can achieve rectification ratios comparable to those of conventional rectifiers.\cite{chen2017molecular} Reproducibility of the molecular junctions continues, nevertheless, to be a problem.

In order to understand the experimentally-observed transport behavior, it is necessary to resort (at least on a qualitative level) to a particular transport theory, many of which have been developed over the last few decades. 
The off-resonant transport regime (where the molecular energy levels lie outside of the bias window) is nowadays almost universally described using the non-interacting Landauer approach,\cite{nitzan2001electron} which includes a use of a transmission coefficient,\cite{engelkes2004length} with the results typically yielding a good match between the observed and theoretically predicted behavior.\cite{lindsay2007molecular,jin2013energy}
Simultaneously, it has been repeatedly demonstrated that this non-interacting approach fails in the resonant regime where the effects of electron-vibrational and electron-electron interactions become important.\cite{secker2011resonant,burzuri2016sequential,thomas2019understanding,fung2019breaking} 
Following the early work of Ulstrup, Kuznetsov and coworkers,\cite{friis1998situ,kuznetsov2000mechanisms,kuznetsov2002mechanisms,zhang2008single} as well as more recent studies by Migliore and Nitzan,\cite{migliore2011nonlinear,migliore2013irreversibility} Marcus theory has also become a popular framework to describe charge transport through molecular junctions at relatively high temperatures. This theory has been successfully applied in the resonant transport regime.\cite{thomas2019understanding,yuan2018transition,jia2016covalently} 
As we shall demonstrate, due to the lack of lifetime broadening in the conventional Marcus theory, it may  fail to correctly describe charge transport in the off-resonant regime. 
Recently, however, a relatively simple theory (which we shall refer to as the generalized theory) unifying the Marcus and Landauer descriptions of charge transport has been developed\cite{sowa2018beyond} (see also Refs.~\onlinecite{sowa2020beyond} and \onlinecite{liu2020generalized}). In the present paper, we modify it so as to include entropic effects in the case of polar solvents. We also provide an intuitive derivation of this theory and apply it to study the transport behavior of molecular junctions.

Besides its perturbative nature (with respect to the molecule-lead interactions), the conventional Marcus theory also treats the vibrational degrees of freedom classically.\cite{marcus1985electron} Consequently, it fails to capture the effects of nuclear tunnelling which can still play an important role in overall charge transport characteristics, even at around room temperature when high frequency vibrations are involved, particularly in the `inverted' region.\cite{barbara1996contemporary,may2008charge} This inverted region has been recently observed experimentally in charge transport through molecular junctions.\cite{yuan2018transition,kang2020interplay}
Therefore, in the last part of this work, we demonstrate how lifetime broadening can be incorporated into the Marcus-Levich-Dogonadze-Jortner-type description of molecular conduction.


\section{Theory \label{model}}
We are interested in molecular junctions comprising a molecular system weakly coupled to two metallic electrodes.
At zero bias the molecular system within the junction is found in the $N$ charge state. As the bias is increased, the charging of the molecule -- populating the $N+1$ (or $N-1$) charge state -- will eventually become possible. For simplicity, we assume that each of the two considered charge states is non-degenerate, and ignore any  excited electronic states.
Then, the molecular system in question can be modelled as a single energy level with energy $\varepsilon_0$ which corresponds to the chemical potential for the charging of the molecular system. We note that, generally, in the presence of electron-vibrational interactions and molecule-lead coupling, the position of the molecular energy level will be renormalized as compared to its gas-phase value. Since, experimentally, the position of the molecular level is typically an empirical parameter, here we simply absorb all these renormalizations into $\varepsilon_0$. 

\begin{figure}[ht]
    \centering
    \includegraphics{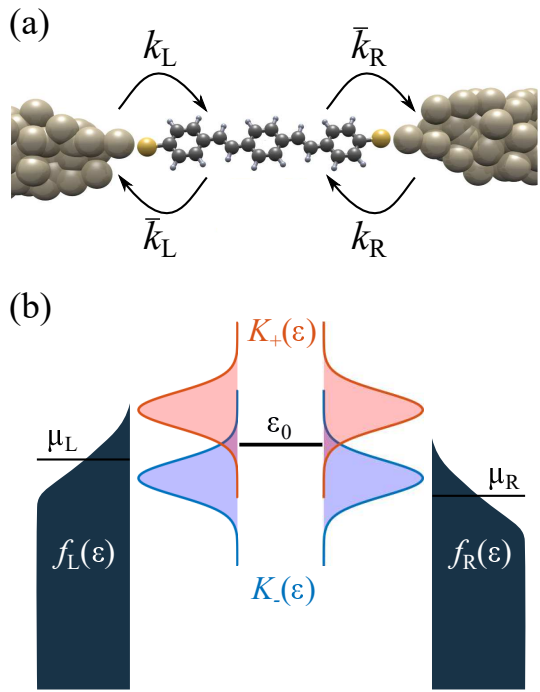}
    \caption{(a) Artistic impression of a single-molecule junction. The effective rates of electron transfer on and off the molecular system are denoted by $k_\mathrm{L}$, $k_\mathrm{R}$ and $\bar{k}_\mathrm{L}$, $\bar{k}_\mathrm{R}$, respectively. (b) Schematic illustration of the rate-equation model considered here; $f_l(\epsilon)$ denotes the Fermi distribution in the lead $l$. $K_\pm(\epsilon)$ are the molecular densities of states. }
    \label{fig1}
\end{figure}
In this work it will be sufficient to model charge transport through the  junction using a rate equation approach. 
As schematically shown in Fig.~\ref{fig1}, charge transport through the weakly-coupled single-molecule junction can be modelled as a series of electron transfers taking place at the left (L) and right (R) electrode. 
In what follows, we will work within the wide-band approximation.\cite{galperin2006resonant,migliore2011nonlinear} We will assume that each of the leads has a constant density of states [$\varrho_l(\epsilon) = \mathrm{const.}$ where $l = \mathrm{L,R}$] and that the electronic coupling between the molecular energy level and a continuum of energy levels in the leads is also constant ($V_l= \mathrm{const.}$ where $V_l$ is the molecule-lead coupling matrix element).

The populations of the $N$ ($P_N$) and $N+1$ ($P_{N+1}$) charge states can be found by considering the following pair of rate equations:
\begin{align}
   \dfrac{\mathrm{d} P_N}{\mathrm{d} t} &= -(k_\mathrm{L} + k_\mathrm{R}) P_N + (\bar{k}_\mathrm{L} + \bar{k}_\mathrm{R}) P_{N+1}~,\\
   \dfrac{\mathrm{d} P_{N+1}}{\mathrm{d} t} &=  - (\bar{k}_\mathrm{L} + \bar{k}_\mathrm{R}) P_{N+1} + (k_\mathrm{L} + k_\mathrm{R}) P_N~,   
\end{align}
where $k_l$ and $\bar{k}_l$ are the rates of electron hopping on and off the molecular structure at the $l$ interface, respectively, as denoted in Fig.~\ref{fig1}.
In the steady-state limit, $\mathrm{d} P_N/{\mathrm{d} t} = {\mathrm{d} P_{N+1}}/{\mathrm{d} t} = 0$, it has the solution
\begin{equation}
    P_N = \dfrac{\bar{k}_\mathrm{L} + \bar{k}_\mathrm{R}}{k_\mathrm{L} + k_\mathrm{R} + \bar{k}_\mathrm{L} + \bar{k}_\mathrm{R}} ~,
\end{equation}
and $P_{N+1} = 1 - P_N$.
The current through the junction can be determined by considering either the left or the right molecule-lead interface. Considering, for instance, the left interface, the current through the junction is given by:
\begin{equation}
    I = e \left[k_\mathrm{L} P_N - \bar{k}_\mathrm{L} P_{N+1}\right]~, 
\end{equation}
which gives the well-known expression:\cite{migliore2011nonlinear,zhang2008single}
\begin{equation}\label{current}
    I = e \dfrac{k_\mathrm{L} \bar{k}_\mathrm{R} - k_\mathrm{R} \bar{k}_\mathrm{L} }{k_\mathrm{L} + k_\mathrm{R} + \bar{k}_\mathrm{L} + \bar{k}_\mathrm{R}}~.
\end{equation}
Although in this work we shall consider a non-degenerate electronic level, the (spin) degeneracy of the electronic level in question can be relatively easily introduced into this model, see for instance Ref.~\cite{thomas2019understanding}. 

The rates of electron transfers  in Eq.~\eqref{current} are given by:\cite{chidsey1991free,sowa2018beyond,gerischer1969charge}
\begin{align}\label{rate1}
    k_l &= \dfrac{2}{\hbar}\Gamma_l \int_{-\infty}^\infty \dfrac{\mathrm{d}\epsilon}{2\pi} f_l(\epsilon) K_+(\epsilon) ~,\\
    \bar{k}_l &= \dfrac{2}{\hbar}\Gamma_l \int_{-\infty}^\infty \dfrac{\mathrm{d}\epsilon}{2\pi} [1-f_l(\epsilon)] K_-(\epsilon) ~,    \label{rate2}
\end{align}
where $f_l(\epsilon) = 1/[\exp((\epsilon - \mu_l)/k_\mathrm{B}T) + 1]$ is the Fermi distribution, $\mu_l$ is the chemical potential of the lead $l$, and $\Gamma_l$ is the  strength of the molecule-lead interaction: 
\begin{equation}\label{gammal}
    \Gamma_l = 2\pi \lvert V_l \rvert^2 \varrho_l~,
\end{equation}
where $\varrho_l$ is the (constant) density of states in the lead $l$ (we make use of this wide-band approximation throughout).
$K_\pm(\epsilon)$ are the molecular densities of states for the relevant processes.
As we shall demonstrate in Section \ref{deriv}, they are given by
\begin{multline}\label{gmm}
    K_\pm (\epsilon) =  \int_{-\infty}^\infty \mathrm{d} E \dfrac{1}{\sqrt{4\pi \lambda k_\mathrm{B} T}}  \times \\ \exp\left(-\dfrac{[\lambda \pm (E - T \Delta S^\circ - \epsilon)]^2}{4\lambda k_\mathrm{B} T}\right)  \dfrac{\Gamma}{(E-\varepsilon_0)^2 + \Gamma^2} ~,    
\end{multline}
where $\lambda$ is the classical reorganisation energy, $\Gamma$ is the lifetime broadening, $\Gamma = (\Gamma_\mathrm{L} + \Gamma_\mathrm{R})/2$, and $\Delta S^\circ$ the entropy change associated with the considered heterogeneous electron transfer  ($\Delta S^\circ$ typically takes negative values when charged species are produced in a polar solvent).
The entropic effects, which will be discussed below, arise from the presence of the $T \Delta S^\circ$ term in Eq.~\eqref{gmm} and, physically, stem predominantly from the changes in the solvent librational-rotational frequencies of the solvent which depend on the charge on the electroactive molecule in the junction (an effect omitted in all `spin-boson' treatments of electron transfer).\cite{marcus1985electron} We note that the entropic effects are therefore not accounted for in descriptions of molecular conduction which treat the nuclear environment quantum-mechanically. It is well-known however that they can play a significant role in electron transfer reactions in polar solvent environments.\cite{marcus1986relation,marcus1985electron,marcus1975electron}

What is the physical meaning of the Eqs.~(\ref{rate1}) and (\ref{rate2})? The overall rate of electron transfer from the lead onto the molecule ($k_l$) can be understood as a sum of the rates for all the possible electron transfers from the continuum of donor states (the population of each of which is determined by the Fermi-Dirac distribution), and conversely for the rate of an electron transfer off the molecular system ($\bar{k}_l$).
$\Gamma_l$ in Eq.~\eqref{gammal}, in units of $\hbar$, is the well-known Golden Rule rate constant for electron transfer from the electronic state of the molecule into the electronic states of the lead, evaluated at the same energy. By microscopic reversibility the rate constant for the isoenergetic reverse step has the same value. 
 
\subsection{Expression for the rate constant \label{deriv}}
In this section, we provide an intuitive derivation for the molecular densities of states $K_\pm(\epsilon)$ from the perspective of the classical theory of electron transfer. For more rigorous derivation we refer the reader to our earlier work in Ref.~\onlinecite{sowa2018beyond} (leading, however, to the omission of the entropic term in Eq.~\eqref{gmm}).

Let us consider a (non-adiabatic) electron transfer between a single band in a metallic lead $l$ (with electrochemical potential $\epsilon$) and the molecular level in question (with energy $\varepsilon_0$). According to the conventional theory of non-adiabatic electron transfer, the rate constant of this process is given by\cite{van1974nonadiabatic,ulstrup1975effect,kestner1974thermal,may2008charge}
\begin{equation} \label{ket}
    k^\mathrm{ET} =  \dfrac{2\pi}{\hbar} \lvert V_l \rvert^2 \ \mathrm{FCWD}~,
\end{equation}
where $V_l$ is the coupling matrix element and FCWD is the Franck-Condon-weighted density of states. 
In what follows, we shall treat the vibrational dynamics classically as it is done within the Marcus theory\cite{marcus1956theory} (although a number of ways to include nuclear tunnelling in a Marcus-type description have been developed\cite{hopfield1974electron,jortner1976temperature,sowa2018beyond}). Later, we will also assume that the nuclear degrees of freedom are thermalized at all times. [We note that methods accounting for non-equilibrium vibrational effects in charge transfer and transport (while treating the vibrational environment classically) have also been developed.\cite{sumi1986dynamical,dou2018broadened,kirchberg2020charge}] In the classical limit, FWCD is therefore given by:\cite{ulstrup1975effect,kestner1974thermal,may2008charge,marcus1985electron}
\begin{equation}
    \mathrm{FCWD} = \dfrac{1}{\sqrt{4\pi\lambda k_\mathrm{B}T}} \exp \left(-\dfrac{[\lambda + (\Delta E - T \Delta S^\circ)]^2}{4\lambda k_\mathrm{B}T} \right) ~,
\end{equation}
where $\lambda$ is the reorganisation energy, and $\Delta E$ and $\Delta S^\circ$ are the energy and entropy differences between the `products' and the `reactants' of the considered process, respectively.

\begin{figure}
    \centering
    \includegraphics{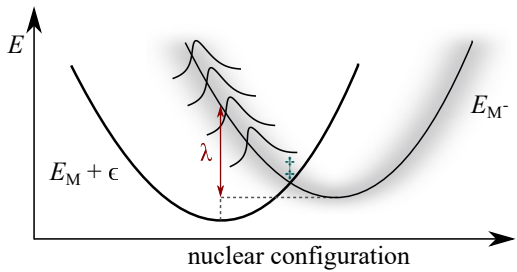}
    \caption{Schematic illustration of the origins of the generalized theory. Parabolas describe the free energies of the reactants and products (of the considered electron transfer) as a function of the nuclear coordinate. Molecule-lead coupling results in broadening of the parabola corresponding to the $N+1$ charge state (M$^{-}$). Note that the shading does not show the dramatic effect that the Lorentzian tails can have on $K_\pm(\epsilon)$.}
    \label{origins}
\end{figure}

Here, we wish to account for the fact that due to the coupling to metallic leads, the state corresponding to the `products' has a finite lifetime (i.e.~is lifetime-broadened, see Fig.~\ref{origins}). 
We therefore assume that the electronic state corresponding to the `products' comprises a continuum of states with the molecular density of states $\rho(E)$ such that:
\begin{equation}
    \int_{-\infty}^\infty \mathrm{d}E \ \rho(E) = 1~.
\end{equation}

Then, the rate of electron transfer (between the single considered metallic band and the molecular energy level) is given by the integral:
\begin{equation} \label{knew}
      k^\mathrm{ET} = \int_{-\infty}^\infty \mathrm{d}E \ \dfrac{2\pi}{\hbar} \ \lvert V_l \rvert^2 \ \mathrm{FCWD}(E) \ \rho(E)~,  
\end{equation}
where the $\mathrm{FCWD}(E)$ is given by:
\begin{equation}
   \mathrm{FCWD}(E) = \dfrac{1}{\sqrt{4\pi\lambda k_\mathrm{B}T}} \exp \left(-\dfrac{[\lambda + (E - T \Delta S^\circ - \epsilon)]^2}{4\lambda k_\mathrm{B}T} \right)~.
\end{equation}

In order to determine $\rho(E)$, let us consider the wavefunction $\psi(t)$ for the molecular energy level. It can be written (in units of $\hbar$) as:
\begin{equation}
    \psi(t) = \theta(t) \left[ \exp(-\mathrm{i} \varepsilon_0 t)\right] \left[ \exp(-\Gamma t)\right] \psi(0) ~,
\end{equation}
where $\theta(t)$ is the Heaviside step function, $\Gamma/\hbar$ is the lifetime (decay constant) for the state in question, and we assume that $\psi(0)$ is normalized.
In the energy space, the corresponding function can be obtained by means of a Fourier transform:
\begin{equation}
    \phi(E) = \int_{-\infty}^\infty \mathrm{d} t  \:\psi(t) \exp(\mathrm{i} E t) = \psi(0) /(\Gamma + \mathrm{i}(E-\varepsilon_0)) ~.
\end{equation}
The probability density $\rho(E)$ is proportional to $\lvert \phi(E)\rvert^2$, i.e.
\begin{equation}
    \rho(E) = C^2  \lvert \phi(E)\rvert^2~,
\end{equation}
where $C$ is the normalisation factor. Since $\lvert \psi(0)\rvert^2 =1$, we obtain
\begin{equation}
   \rho(E) =  \dfrac{\Gamma}{ \pi[(E-\varepsilon_0)^2 + \Gamma^2]}~.
\end{equation}

The electron-transfer rate given in Eq.~\eqref{knew} therefore becomes 
\begin{multline} \label{kkk}
      k^\mathrm{ET} = \int_{-\infty}^\infty \mathrm{d}E \ \dfrac{2\pi}{\hbar} \ \lvert V_l \rvert^2 \ \dfrac{1}{\sqrt{4\pi\lambda k_\mathrm{B}T}} \times \\ \exp \left(-\dfrac{[\lambda + (E - T \Delta S^\circ - \epsilon)]^2}{4\lambda k_\mathrm{B}T} \right) \dfrac{\Gamma}{\pi[(E-\varepsilon_0)^2 + \Gamma^2]}~.    
\end{multline}
The overall (effective) rate of electron transfer from the metallic electrode and onto the molecular level (or \textit{vice versa}) is simply a sum of the rates of individual electron transfers weighted by the Fermi distribution and the lead density of states, as described by Eqs.~\eqref{rate1} and \eqref{rate2}. From Eq.~\eqref{kkk} we therefore obtain the expression for $K_\pm(\epsilon)$ given in Eq.~\eqref{gmm}.
This result constitutes the basis of what we will refer to as a generalized theory (which shall be discussed in greater detail in Section \ref{gmarcus}).

\subsection{Landauer and Marcus limits \label{limitss}}
In this section we will demonstrate that the conventional Landauer and Marcus theories can be obtained as the limiting cases of the generalized theory.
As can be seen in Eq.~\eqref{gmm}, $K_\pm(\epsilon)$ in the generalized theory is given by a convolution of the Lorentzian and Gaussian profiles.
Let us first consider the case of vanishing reorganisation energy. Then, $\sqrt{4\lambda k_\mathrm{B}T}/\Gamma \rightarrow 0$, i.e.~the Gaussian profile in Eq.~\eqref{gmm} becomes very narrow as compared to the Lorentzian.
We also know that with vanishing reorganization energy the $-T \Delta S^\circ$ term also vanishes since the time is too short for the changes in polar solvent configurations to contribute.
In this limit, therefore, the relevant Gaussian function becomes
\begin{equation}
 \dfrac{1}{\sqrt{4\pi \lambda k_\mathrm{B} T}} \exp\left(-\dfrac{[\lambda \pm (E - T \Delta S^\circ - \epsilon)]^2}{4\lambda k_\mathrm{B} T}\right) \rightarrow \delta(E-\epsilon)  ~,
\end{equation}
and Eq.~\eqref{gmm} simplifies to
\begin{equation} \label{lorentz}
    K_\pm(\epsilon) = \dfrac{\Gamma}{(\epsilon - \varepsilon_0)^2 + \Gamma^2}~,
\end{equation}
where the molecular densities of states are identical for an electron transfer on and off the molecular system (microscopic reversibility). Inserting Eq.~\eqref{lorentz} into Eq.~\eqref{current} allows us to reduce the expression for electric current to the usual Landauer (Landauer-B\"uttiker) approach.\cite{landauer1957spatial,imry1999conductance,zimbovskaya2013transport,nitzan2006chemical,esposito2009transport} It becomes:
\begin{equation} \label{LB}
    I = \dfrac{e}{\hbar} \int_{-\infty}^\infty \dfrac{\mathrm{d}\epsilon}{2\pi} \left( f_\mathrm{L}(\epsilon) - f_\mathrm{R}(\epsilon) \right) \mathcal{T}(\epsilon) ~,
\end{equation}
where $\mathcal{T}(\epsilon)$ is the transmission function, here given by a Breit-Wigner resonance:\cite{breit1936capture}
\begin{equation}
    \mathcal{T}(\epsilon) = \dfrac{\Gamma_\mathrm{L} \Gamma_\mathrm{R}}{(\epsilon - \varepsilon_0)^2 + \Gamma^2}~.
\end{equation}
Furthermore, it is  instructive to consider the Landauer approach in the limit of zero temperature and for a constant transmission function $\mathcal{T}(\epsilon) =\mathcal{T}$. Then, Eq.~\eqref{LB} becomes:
\begin{equation}
   I = \dfrac{e}{h} (\mu_\mathrm{L} - \mu_\mathrm{R}) \mathcal{T} = \dfrac{e^2}{h} V_\mathrm{b} \mathcal{T} ~. 
\end{equation}
Introducing an additional factor of two to account for the spin degeneracy of the considered level, we recover the celebrated Landauer formula for the electronic conductance:
\begin{equation}\label{glan}
    G = \dfrac{\mathrm{d}I}{\mathrm{d}V_\mathrm{b}} = \dfrac{2e^2}{h} \mathcal{T}~,
\end{equation}
where $\mathcal{T}$ can vary between 0 and 1.\cite{landauer1957spatial,landauer1989conductance}
For completeness, an alternative derivation of Eq.~\eqref{glan} is given in the Appendix \ref{appL}.

Next, we consider Eq.~\eqref{gmm} in the limit when $\Gamma/\sqrt{4\lambda k_\mathrm{B}T} \rightarrow 0$, that is when the width of the Lorentzian profile is negligible compared to that of the Gaussian profile. Then,
\begin{equation}
  \dfrac{\Gamma}{(E-\varepsilon_0)^2 + \Gamma^2} \rightarrow \pi \: \delta(E-\varepsilon_0)~, 
\end{equation}
and $K_\pm(\epsilon)$ in Eq.~\eqref{gmm} take the familiar form:
\begin{equation} \label{marcus}
    K_\pm(\epsilon) = \sqrt{\dfrac{\pi}{4\lambda k_{\mathrm{B}} T}} \exp\left( -\dfrac{[\lambda \pm ( \varepsilon_0 - T \Delta S^\circ - \epsilon)]^2}{4\lambda k_{\mathrm{B}} T}\right)  ~. 
\end{equation}
Together with Eqs.~\eqref{rate1} and \eqref{rate2}, Eq.~\eqref{marcus} constitutes Marcus (Marcus-Levich-Dogonadze-Hush-Chidsey-Gerischer) theory of transport.\cite{marcus1985electron,marcus1956theory,chidsey1991free,gosavi2000nonadiabatic,migliore2012relationship,migliore2011nonlinear} 

As was previously discussed,\cite{migliore2012relationship} Landauer and Marcus theories describe the opposite limits of charge transport mechanism. The former describes transport as a coherent process. 
In the latter, meanwhile, it is assumed that before and following an electron transfer (from one of the metallic leads) the vibrational environment relaxes and the charge density localizes on the molecular system (until it tunnels out into the metallic lead).\footnote{It is interesting to note that a Gaussian profile is sometimes introduced \textit{ad hoc} into the Landauer framework in order to explain the experimentally-observed behavior.\cite{chen2017molecular}}

\subsection{Back to the generalized approach \label{gmarcus}}
Here, we return to the generalized theory derived in Section \ref{deriv} which, as we have shown above, unifies the conventional Marcus and Landauer theories of molecular conduction.
(We note that the performance of our generalized theory is yet to be validated in the intermediate regime, between the Landauer and Marcus limits, by a detailed comparison with exact quantum-mechanical calculation or experiment.)
As can be clearly seen in Eq.~\eqref{gmm}, the molecular densities of states $K_\pm(\epsilon)$ in the generalized theory are given by a Voigt function (a convolution of a Gaussian and a Lorentzian).\cite{armstrong1967spectrum}

It is instructive to consider Eq.~\eqref{gmm} far away from resonance, i.e.~when $\lvert \epsilon- \varepsilon_0 \rvert \gg \lambda, k_\mathrm{B}T, \Gamma$. 
In this limit, the Lorentzian and Gaussian profiles in Eq.~\eqref{gmm} are centered very far apart from each other (on the $E$-axis) so that the wings of the Lorentzian are virtually constant over the width of the Gaussian profile.  Consequently, the integral in Eq.~\eqref{gmm} returns simply the value of the Lorentzian profile (far away from the resonance).
Therefore, as we also more rigorously show in Appendix \ref{appA}, far away from resonance the $K_\pm(\epsilon)$ in Eq.~\eqref{gmm} can be approximated as:
\begin{equation} \label{approxmm}
    K_\pm(\epsilon) \approx \dfrac{\Gamma}{(\epsilon- \varepsilon_0)^2}~.
\end{equation}
This is a significant result for several reasons. Firstly, we note  that far from resonance $K_\pm (\epsilon)$ are independent of temperature. Furthermore, in the limit of $\lvert \epsilon- \varepsilon_0 \rvert \gg \Gamma$, the same expression can be obtained from the Landauer expression for $K_\pm(\epsilon)$ given in Eq.~\eqref{lorentz}.
Therefore, the generalized theory coincides with the Landauer approach not only for vanishing reorganisation energy (as we have previously discussed) but also far away from resonance: 
in the deep off-resonant regime an interacting system can be approximated as a non-interacting one.
This result is in agreement with a multitude of experimental studies which, as discussed above,  successfully modelled off-resonant transport using the Landauer approach.
Off-resonant charge transport is often the mechanism of conduction through molecular junctions especially at relatively low bias voltage and it is possible that it may also account for the long-range electron transport observed through DNA-based systems.\cite{beratan2017charge,kim2016intermediate,wierzbinski2013single,dauphin2019high}

In our previous work, we have studied the $IV$ characteristics and the thermoelectric response predicted by the generalized theory.\cite{sowa2018beyond,sowa2019marcus} 
Here, we will explore the temperature-dependence of electric current predicted by this approach in various transport regimes, and compare it to that predicted by the conventional Landauer and Marcus approaches.

\begin{figure}
    \centering
    \includegraphics{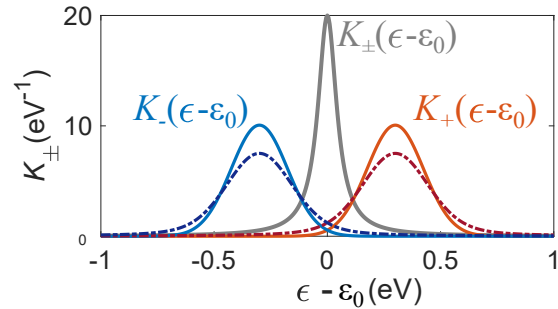}
    \caption{Molecular densities of states $K_\pm(\epsilon)$ as present in the (i) Landauer approach [solid thick line], (ii) Marcus theory [solid lines], and (iii) generalized theory [dashed lines]. $K_\pm(\epsilon)$ were calculated for instructive values of $\lambda = 0.3$ eV and $\Gamma = 50$ meV at $T=300$ K. For simplicity, we also set $\Delta S^\circ = 0$.}
    \label{fig2}
\end{figure}
\subsection{Some general remarks}
In summary, charge transport through a weakly-coupled molecular junction (modelled as a single electronic level) can be described as a series of electron transfers with the molecular densities of states taking a form of a Lorentzian (Landauer approach), Gaussian (Marcus theory), and Voigt functions (generalized theory), as in Fig.~\ref{fig2}.
We note that within all of these approaches\footnote{Naturally, this does not hold for the approximation of $K_\pm(\epsilon)$ given in Eq.~\eqref{approxmm} which is valid only on the part of the energy domain.}
\begin{equation} \label{norma}
    2\int_{-\infty}^\infty \dfrac{\mathrm{d}\epsilon}{2\pi} K_\pm(\epsilon) = 1~,
\end{equation}
so that at very high bias $k_\mathrm{L} = \Gamma_\mathrm{L}/\hbar$, $\bar{k}_\mathrm{R} = \Gamma_\mathrm{R}/\hbar$, and $k_\mathrm{R}=\bar{k}_\mathrm{L} = 0$, or \textit{vice versa}.
Therefore, in the limit of very high bias voltage,   we obtain the well-known value of electric current:
\begin{equation}
    I = \dfrac{e}{\hbar} \: \dfrac{\Gamma_\mathrm{L}\Gamma_\mathrm{R}}{\Gamma_\mathrm{L}+\Gamma_\mathrm{R}}~,
\end{equation}
which is independent of the chosen theoretical approach (and so also of the strength of the vibrational coupling).

We again stress that all the theories discussed here  assume the presence of only a single molecular electronic energy level (in each of the two considered charge states). They are therefore valid (in their presented form) at sufficiently low bias voltages such that the excited electronic states can be disregarded, and far away from the remaining charge degeneracy points (where populating charge states other than $N$ and $N+1$ becomes possible).

\subsection{Single-barrier model}
In the above, the molecular system within the junction was effectively modelled as a well potential with two tunnelling barriers -- one at each of the molecule-lead interfaces.
It is also worth to mention another relatively simple theoretical model which is somewhat complementary to what has been discussed here. Namely, it is possible to approximately model the molecular junction as a single (typically trapezoidal) tunnelling barrier,\cite{choi2008electrical,beebe2006transition, wang2003mechanism, wold2001fabrication} and obtain the current-voltage characteristics using the Simmons model.\cite{simmons1963generalized} Within this approach, no additional charge density can localise on the molecule.
It does not therefore account for the reorganisation of the vibrational environment associated with the charging of the molecule in the junction and is typically justified only in a deep off-resonant regime.
This approach has been successfully used to account for the observed charged transport through molecular system with high-lying molecular energy levels.\cite{choi2008electrical,beebe2006transition, wang2003mechanism, wold2001fabrication}

\section{Comparison of the conduction theories \label{3a}}
In this section, we explore the temperature-dependence of the electric current as predicted by the three approaches described above. 
We first calculate the $IV$ characteristics for the energy level lying at $\varepsilon_0 = 0.5$ eV above the Fermi levels of the unbiased leads. Where appropriate, we set $\lambda = 0.3$ eV (c.f.~Ref.~\cite{thomas2019understanding}), assume relatively weak and symmetric molecule-lead coupling: $\Gamma_\mathrm{L} = \Gamma_\mathrm{R} = 1$ meV and, for simplicity, set $T \Delta S^\circ = 0$. Experimentally, values of lifetime broadening from less than 1 $\mu$eV up to a few hundred meV have been observed.\cite{frisenda2016transition,thomas2019understanding,fung2019breaking,capozzi2015single} This large spread in the observed $\Gamma$ stems most likely from variations in the nature of molecule-lead contacts (the electronic coupling is typically assumed to decay exponentially with distance) as well as in the densities of states in the metallic electrodes (which depend on the exact atomic structure of the metallic tips). 
The chemical potentials of the leads are determined by the applied bias voltage $V_\mathrm{b}$: $\mu_\mathrm{L} = -\lvert e\rvert \alpha V_\mathrm{b}$ and $\mu_\mathrm{R} = \lvert e\rvert (1-\alpha) V_\mathrm{b}$.
The parameter $\alpha$ accounts for how the potential difference is distributed between the left and right electrodes (and varies between 0 and 1), see Ref.~\cite{datta1997current} for a detailed discussion. In particular if $\alpha = 0.5$, the bias voltage is applied symmetrically resulting in a symmetric $IV$ curve. Otherwise, the bias is applied asymmetrically giving rise to current rectification (asymmetrical $IV$ characteristics).\cite{chen2017molecular,capozzi2015single}

\begin{figure}[h]
    \centering
    \includegraphics{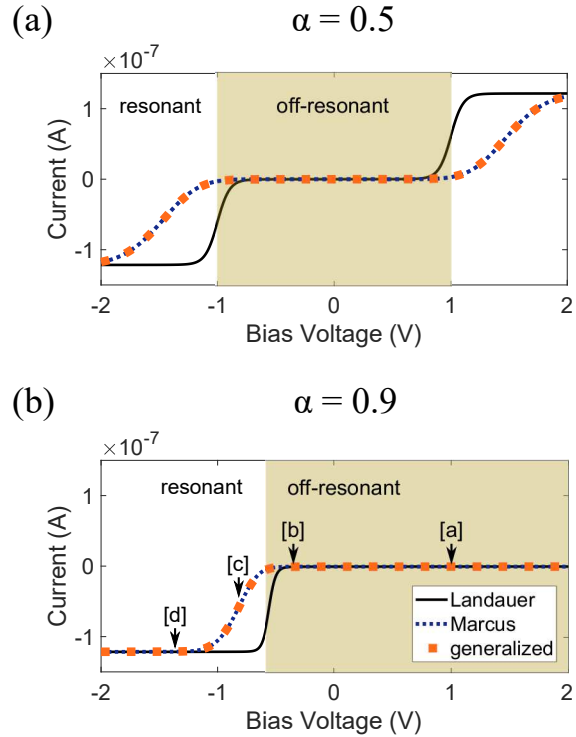}
    \caption{$IV$ characteristics calculated using the Landauer, Marcus and generalized approaches for: (a) $\alpha = 0.5$, and (b) $\alpha = 0.9$. We set the position of the molecular level above the Fermi level of the unbiased leads $\varepsilon_0 = 0.5$ eV, $\Gamma_\mathrm{L} = \Gamma_\mathrm{R} = 1$ meV, $\lambda = 0.3$ eV (in Marcus and generalized approaches), and $T=300$ K. The shaded area marks the off-resonant regime (when the molecular level lies outside of the bias window). Note that Marcus and generalized theory curves appear to closely overlap in the resonant transport regime. }
    \label{IVs}
\end{figure}

We begin by calculating the $IV$ characteristics for $\alpha = 0.5$ and $\alpha =0.9$ in Fig.~\ref{IVs}(a) and (b), respectively. All of them exhibit the expected behavior (for a single-level model): region of suppressed current at low bias voltage (where the molecular energy level is found outside of the bias window) followed by a rise in current and an eventual plateau in the deep resonant regime.
In the presence of electron-vibrational interactions (i.e.~within the Marcus and generalized approaches), we can observe lower values of current as the molecular energy level enters the bias window. This is fundamentally an example of a Franck-Condon blockade.\cite{koch2005franck,bevan2018relating}
Furthermore, due to relatively small $\Gamma$, the Marcus and generalized theory predict seemingly very similar behavior. As we shall demonstrate (\textit{vide infra}), the differences between these approaches become appreciable in the off-resonant transport regime.

\begin{figure*}[ht]
    \centering
    \includegraphics{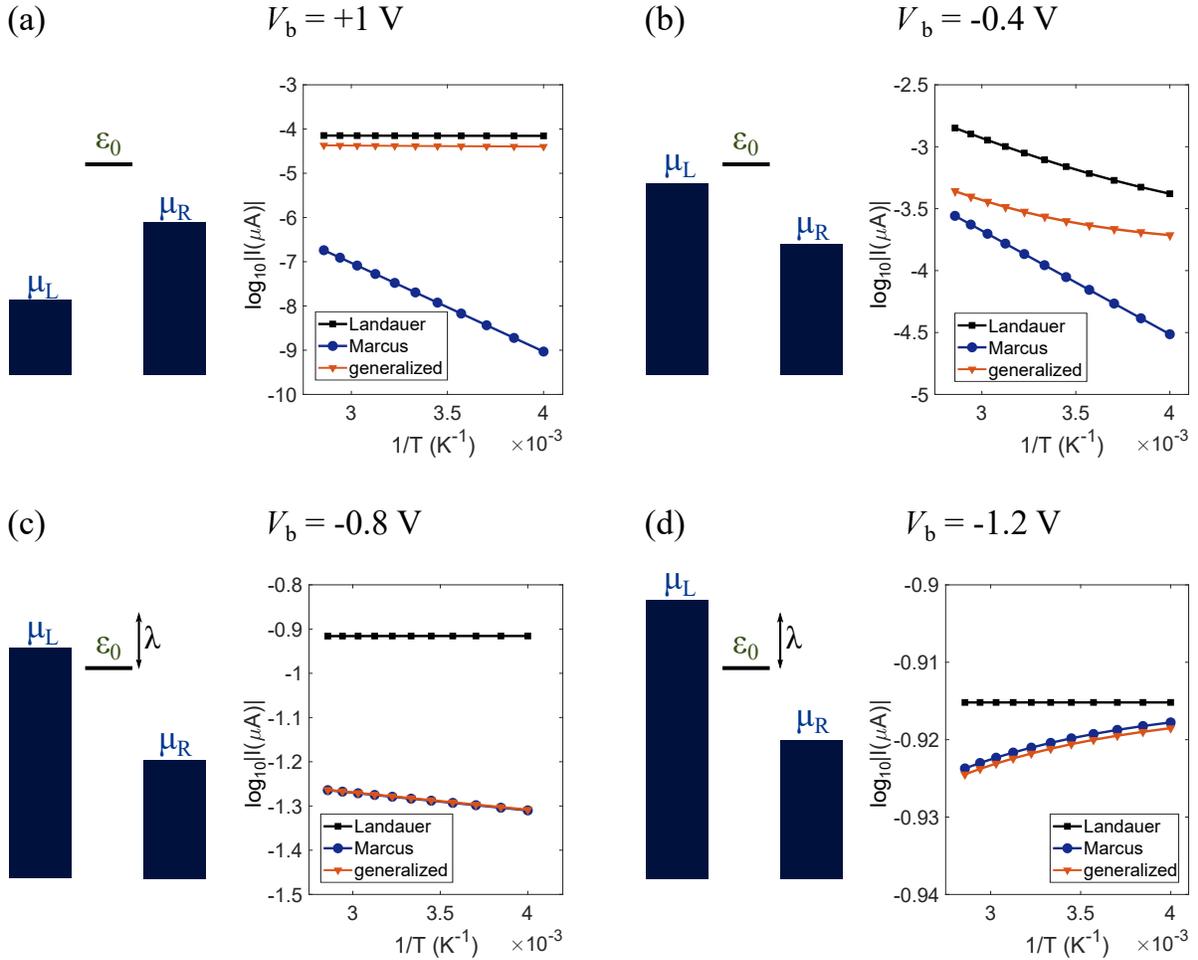}
    \caption{Arrhenius plots of electric current [$\mathrm{log}_{10}(I)$ \textit{vs.}~$1/T$] at bias voltage $V_\mathrm{b}=\{1,-0.4,-0.8,-1.2\}$ V as a function of temperature. Other parameters as in Fig.~\ref{IVs}(b): $\alpha =0.9$, $\Gamma_\mathrm{L} = \Gamma_\mathrm{R} = 1$ meV, $\lambda = 0.3$ eV.  Left panels  schematically show the relative positions of the molecular energy level and the chemical potentials of the leads (for clarity, broadening of the Fermi distributions in the leads is not shown).}
    \label{temp_dep}
\end{figure*}

We now turn to examine the temperature dependence of the electric current as predicted by the three approaches considered here.
This is done in Fig.~\ref{temp_dep} which shows the electric current as a function of temperature (on an Arrhenius plot) for different values of the bias voltage. We consider current at four different bias voltages [as marked by arrows in Fig.~\ref{IVs}(b)], initially disregarding the entropic effects ($\Delta S^o = 0$).

Within the Landauer approach, the temperature dependence of the electric current stems solely from the temperature dependence of the Fermi distributions in the leads. Consequently, the electric current is almost independent of temperature when the molecular energy level lies far away from the bias window [Fig.~\ref{temp_dep}(a)], increases with temperature in the case of near-resonant transport [Fig.~\ref{temp_dep}(b)], and decreases with increasing temperature in the resonant transport regime [Figs.~\ref{temp_dep}(c) and (d)] although this effect can be relatively modest.

In contrast, within the Marcus approach, the temperature-dependence is determined by both the temperature dependence of the Fermi distributions in the leads and that of the Marcus rates in Eq.~\eqref{marcus}.
The latter contribution typically dominates and usually exhibits an exponential dependence on inverse temperature. Indeed, we observe an Arrhenius-type behavior in the far off-resonant scenario [Fig.~\ref{temp_dep}(a)]: electric current depends exponentially on inverse temperature and is greatly suppressed, as compared to that predicted by the Landauer theory. The same is true in the near-resonant case [Fig.~\ref{temp_dep}(b)].
In the resonant regime, the electric current increases (in an Arrhenius-type fashion) with temperature as long as the chemical potential of the left lead satisfies $\mu_\mathrm{L} < \varepsilon_0 + \lambda$ [Fig.~\ref{temp_dep}(c)].\cite{migliore2012relationship}
In the deep resonant regime (for $\mu_\mathrm{L} > \varepsilon_0 + \lambda$), broadening of both the Fermi distributions in the leads and the molecular densities of states $K_\pm(\epsilon)$ leads to a modest decrease in current with increasing temperature  [Fig.~\ref{temp_dep}(d)].

Finally, we consider the generalized theory. Within this approach, the temperature dependence of electric current once again stems from the broadening of the Fermi distributions as well as temperature dependence of the electron transfer rates. The temperature dependence of $K_\pm(\epsilon)$ given in Eq.~\eqref{gmm} is, however, rather non-trivial. 
In the deep off-resonant regime [Fig.~\ref{temp_dep}(a)], electric current is virtually independent of temperature and takes values similar to those predicted by the Landauer approach, see discussion in Section \ref{gmarcus}.
In the near-resonant case  [Fig.~\ref{temp_dep}(b)], electric current generally increases with increasing temperature although in a non-linear fashion different from what is predicted by both the Landauer and Marcus transport theories.
Conversely, in the resonant regime the predictions of the generalized theory closely coincide with those of the conventional Marcus theory.

These results illustrate the fact that both the Landauer and Marcus theories can be used to describe charge transport through molecular junctions in their respective regimes of applicability. As discussed above, these different regimes may even correspond to different ranges of bias voltage for the same molecular junction. 

\section{Entropic effects}
\begin{figure}
    \centering
    \includegraphics{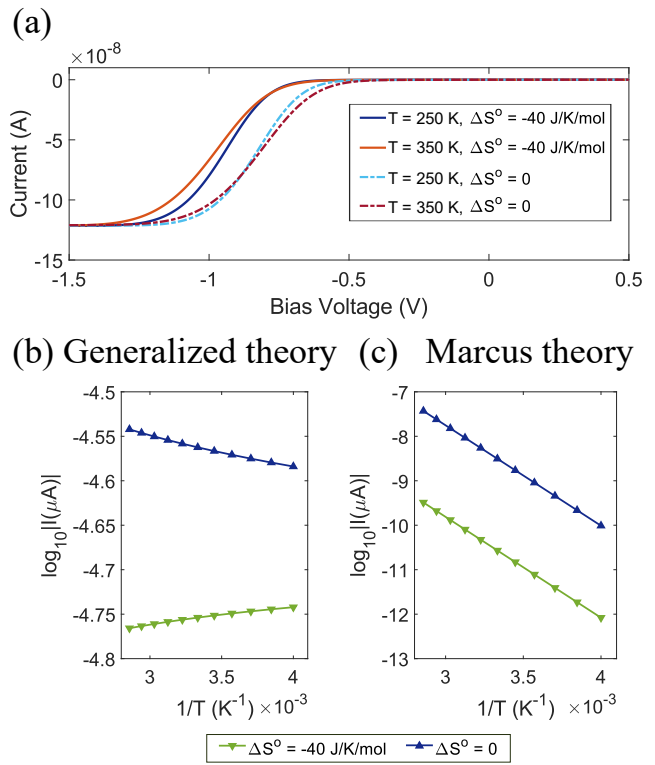}
    \caption{(a) $IV$ characteristics calculated at different temperatures for $\Delta S^o = -40$ and $0$ J K$^{-1}$ mol$^{-1}$. Other parameters as in Fig.~\ref{temp_dep}. (b, c) Temperature dependence of the electric current at $V_b = +0.5$ V calculated using the (b) generalized and (c) conventional Marcus theory.}
    \label{fentropy}
\end{figure}

We next investigate the role of entropic effects in molecular conduction. In accordance with previous experimental studies of electron transfer in polar solvents,\cite{marrosu1990reaction,komaguchi1991entropy,svaan1984temperature} we set $\Delta S^o = -40$ J K$^{-1}$ mol$^{-1}$ (which corresponds to roughly -0.41 meV K$^{-1}$) unless stated otherwise. 
First, in Fig.~\ref{fentropy}(a), using our generalized theory, we calculate the $IV$ characteristics obtained for  $\Delta S^o = 0$ and $-40$ J K$^{-1}$ mol$^{-1}$ and at different temperatures. The current steps, present in the $IV$ characteristics when the molecular energy level falls into the bias window, are significantly shifted for non-zero $\Delta S^o$. 
Furthermore, in the presence of entropic effects, the magnitudes of those shifts are increasing with temperature, while for $\Delta S^o = 0$ increasing temperature leads solely to the broadening of the $IV$ characteristics.
In the resonant (high-current) region, qualitatively identical behavior is also predicted by the conventional Marcus theory (not shown).

The origin of both of these effects can be understood using Eq.~\eqref{gmm}: the inclusion of entropic effects corresponds to an effective (and temperature-dependent) renormalization of the position of the molecular energy level. For negative $\Delta S^o$, this results in a shift of the current step toward higher values of bias voltage (shift in the opposite direction will be observed in the case of transport through a level found below the Fermi level of the unbiased leads). 
From Eqs.~\eqref{gmm} and \eqref{marcus}, it can be inferred that strong entropic effects should be expected when $\lambda + (\varepsilon_0 - \epsilon) =0$. 
It can be indeed seen in Fig.~\ref{fentropy}(a) that inclusion of negative $\Delta S^o$ leads to a negative temperature coefficient of the current (decreasing current with increasing temperature) in the resonant regime. Analogous negative temperature coefficient has been seen experimentally in charge recombination electron transfer reactions in polar liquids when the intrinsic barrier to reaction is small and has been discussed in the literature.\cite{marcus1985electron,marcus1975electron}  The decrease of electric current with increasing temperature can occur when $\Delta S^o$ is negative and the molecular energy level is found above the Fermi levels of the unbiased leads or when $\Delta S^o$ is positive and the and the molecular energy level is found below the Fermi levels of the electrodes.
We also note that the qualitative behavior of the electric current in the resonant regime (as a function of temperature) could be used to experimentally determine the sign of $\Delta S^o$.

In Figs.~\ref{fentropy}(b) and (c), we further consider the temperature dependence of the electric current in the off-resonant regime using the generalized and conventional Marcus theory, respectively (we do not consider here the Landauer theory since it disregards the environmental interactions). In the presence of negative $\Delta S^o$, we observe lower values of electric current through the junction (once again due to the temperature-dependent shift of the effective position of the molecular level).
The electric current predicted by the generalized theory [Fig.~\ref{fentropy}(b)] exhibits only fairly weak temperature dependence, in accordance with previous discussion. In the case of non-zero  $\Delta S^o$, the current very unusually decreases with increasing $T$ as the temperature dependence is dominated by the entropic effect. This can again be explained by the effective renormalization of the position of the molecular level by the entropic term.
On the other hand, within the conventional Marcus theory [Fig.~\ref{fentropy}(c)], we once again observe Arrhenius-type characteristics. Unlike the magnitude of the current, its temperature-dependent behavior is not significantly affected by the entropic effects.
 
In summary, entropic effects (of a realistic magnitude) can result in an unusual temperature-dependent behavior of the electric current. Negative temperature coefficient in particular may serve as an indication of this phenomenon in experimental studies on solvated molecular junctions. 

\section{Marcus-Levich-Dogonadze-Jortner description \label{jortner}}
Thus far, the entire vibrational environment was treated classically.
It is well-known, however, that the high-temperature assumption of Marcus theory is generally not valid at around room temperature for the high-frequency molecular modes. These modes should be treated quantum-mechanically in order to obtain a qualitative agreement with the experimental studies.\cite{miller1984effect,closs1986distance} This need motivated Jortner and coworkers to develop an extension of the classical Marcus theory, known as the Marcus-Levich-Dogonadze-Jortner theory.\cite{jortner1976temperature,ulstrup1975effect} Within this approach, molecular vibrational environment is divided into two components: the low-frequency part typically associated with the outer-sphere environment, and the high frequency part represented by a single effective mode of frequency $\omega_0$. This effective high-frequency mode typically represents molecular vibrational modes corresponding to carbon-carbon and carbon-oxygen double-bond stretches (ubiquitous to most organic structures) and has a frequency of roughly 190 meV ($\sim 1500$ cm$^{-1}$). Then, the rate of electron transfer is given by Eq.~\eqref{ket} with the Franck-Condon-weighted density of states\cite{jortner1976temperature}
\begin{multline}\label{fcwdj}
    \mathrm{FCWD} = \dfrac{1}{\sqrt{4\pi \lambda_\mathrm{out} k_\mathrm{B} T}}\sum_{m=0}^\infty e^{-D} \dfrac{D^m}{m!} \\ \exp\left(- \dfrac{[\Delta E  - T \Delta S^\circ + \lambda_\mathrm{out} + m \omega_0]^2}{4\lambda_\mathrm{out} k_\mathrm{B} T}\right)~,
\end{multline}
where  $\lambda_\mathrm{out}$ is the outer-sphere reorganisation energy. $D$ is the Huang-Rhys parameter for the coupling to the effective high-frequency vibrational mode
\begin{equation}
   D = \dfrac{\lambda_\mathrm{in}}{\omega_0}~,
\end{equation}
where $\lambda_\mathrm{in}$ is the corresponding reorganisation energy.
Marcus-Levich-Dogonadze-Jortner theory (in its original formulation as well as its multi-mode extension) has become the most commonly used way to introduce nuclear tunnelling into the description of electron transfer.\cite{barbara1996contemporary}
We recall that in the conventional Levich-Dogonadze and all similar quantum mechanical treatments the medium in which the charges exist do not contain a $\Delta S^o$ term because of the assumptions tacitly made in treating the environment quantum mechanically.

It is also possible to adapt this theory in the transport setting considered here and incorporate lifetime broadening into this framework.
Using Eq.~\eqref{knew} and the $\mathrm{FCWD}$ factor given in Eq.~\eqref{fcwdj}, the relevant densities of states are given by:
\begin{multline} \label{jortnerlife}
    K_\pm(\epsilon) = \sqrt{\dfrac{\pi}{4 \lambda_\mathrm{out} k_\mathrm{B}T}}\sum_{m=0}^\infty e^{-D} \dfrac{D^m}{m!}  \times  \int_{-\infty}^\infty \mathrm{d} E  \\ \exp\left(-\dfrac{[(\lambda_\mathrm{out} + m \omega_0) \pm (E - T \Delta S^\circ - \epsilon)]^2}{4\lambda_\mathrm{out} k_\mathrm{B} T}\right)  \times \dfrac{\Gamma}{(E - \varepsilon_0)^2 + \Gamma^2}   ~.    
\end{multline}
This constitutes what we shall refer to as the generalized Marcus-Levich-Dogonadze-Jortner (gMLDJ) theory.

\begin{figure}
    \centering
    \includegraphics{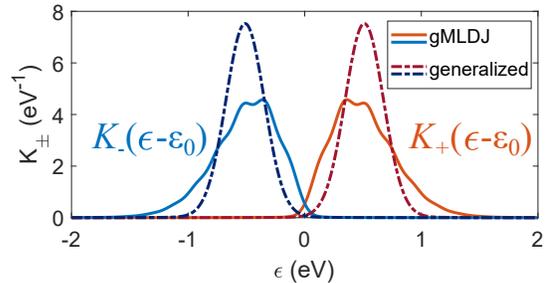}
    \caption{Molecular densities of states $K_\pm(\epsilon)$ calculated for $\Gamma = 5$ meV, $\lambda_\mathrm{out} = 150$ meV, $D = 1.9$, $\omega_0 = 190$ meV (gMLDJ), and $\lambda = \lambda_\mathrm{out} + D \omega_0$ (generalized theory) at $T=300$ K. For simplicity, $\Delta S^\circ = 0$.}
    \label{fig4}
\end{figure}
In Fig.~\ref{fig4}, we plot the molecular densities of states $K_\pm(\epsilon)$ obtained using the generalized Marcus and generalized MLDJ approaches. The latter clearly shows a set of equidistant peaks separated by $\omega_0$ which correspond to the excitations of the (effective) high-frequency  molecular mode.
Since this  high-frequency vibrational mode constitutes a somewhat phenomenological description of the inner-sphere environment (which in reality comprises a set of vibrational modes), the presence of these equally-spaced conductance peaks is an artefact of the Marcus-Levich-Dogonadze-Jortner approach.
Furthermore, the Marcus-Levich-Dogonadze-Jortner approach predicts a much larger magnitude of $K_\pm(\epsilon)$ (as compared to the classical Marcus rates) for both smaller and larger values of $\lvert\epsilon -\varepsilon_0\rvert$, a direct result of incorporating nuclear tunnelling in the Marcus-Levich-Dogonadze-Jortner theory.
All these aspects of Marcus-Levich-Dogonadze-Jortner theory have long been well-understood.\cite{barbara1996contemporary} We note that nuclear tunneling is much more important in the inverted regime than in the normal regime.

In an analogy to what was discussed in Section \ref{limitss}, by setting $\lambda_\mathrm{out} = \lambda_\mathrm{in} = 0$ in the gMLDJ theory we again recover the Landauer description of transport.
Once again, lifetime broadening again becomes especially relevant  in off-resonant regime of transport.
Qualitatively, the behavior which is predicted by this approach in the off-resonant regime will coincide with that of the generalized theory:
inclusion of lifetime broadening will result in increased electric current and its very weak temperature dependence, \textit{c.f.} Section \ref{3a}. 
Finally, we note that lifetime broadening can also be introduced in the multi-mode extension of Marcus-Levich-Dogonadze-Jortner theory (where it would normally be necessary to calculate the Huang-Rhys factor for each of the molecular modes).\cite{ulstrup1975effect,jortner1976temperature} 
This modification would lead, however, to an even more complicated expression and we see little advantage in using such an approach  in practical applications (as opposed to, for instance, the generalized-quantum-master-equation result of Ref.~\onlinecite{sowa2018beyond}).

\section{Concluding Remarks \label{end}}
In this work, we first focused on the recently-derived generalized theory.
We have presented an intuitive derivation of this approach, showed how entropic effects can be incorporated into that formalism, and demonstrated how the conventional Landauer and Marcus approaches can be obtained as limiting cases of this more general approach.
We have further demonstrated that (for relatively weak molecule-lead coupling) the predictions of the generalized  theory coincide very well  with those of Landauer and Marcus theories in the off-resonant and resonant regime, respectively.
Consequently, we believe that the generalized theory correctly describes transport properties of molecular junctions across the entire experimentally-accessible domain (i.e.~in both the resonant and off-resonant regime; provided the high-temperature assumption of Marcus theory is justified).
We have also studied the influence and identified experimental signatures of entropic effects in the molecular electronic conduction in different transport regimes.
Finally, in Section \ref{jortner}, we have shown how lifetime broadening can be introduced into Marcus-Levich-Dogonadze-Jortner theory.
The theory presented here can be also extended beyond the single-level model and thus introduce lifetime-broadening effects into the rate-equation descriptions\cite{migliore2011nonlinear} of multi-level molecular junctions.\footnote{However, an \textit{ad hoc} replacement of the molecule-lead hopping rates (in the conventional rate equation or quantum master equation approaches) with the expressions developed here will yield a theory that does not recover the exact Landauer result for molecular systems with more than one site.} 
Our hope is that this work will inspire a wide use of the theory described here in experimental studies on molecular junctions as well as stimulate empirical exploration of entropic effects in these systems.  


\begin{acknowledgements}
JKS thanks Hertford College, Oxford for financial support, and L. MacGregor for carefully reading the manuscript.
RAM thanks the Office of the Naval Research and the Army Research Office for their support of this research.
\end{acknowledgements}

\section*{Data Availability Statement}
The data that supports the findings of this study are available within the article itself.

\appendix

\section{Alternative derivation of the Landauer formula \label{appL}}
To derive the Landauer formula for the electronic conductance, let us consider a one-dimensional wire connecting two electronic reservoirs with electrochemical potentials $\mu_\mathrm{L}$ and $\mu_\mathrm{R}$, respectively.
If the length of the wire is given by $L$, then the electric current through the considered system is (at zero temperature)  given by:
\begin{equation}
    I = e \: n^+\: \dfrac{v}{L} ~,
\end{equation}
where $v$ is the velocity of the charge carriers within the wire and $n^+$ is the number of states for electrons propagating from left to right within the bias window (i.e.~between $\mu_\mathrm{L}$ and $\mu_\mathrm{R}$).
Assuming that the wire possess a (quasi-)continuum of states, we can write
\begin{equation}\label{I1}
    I = e \: (\mu_\mathrm{L} - \mu_\mathrm{R})  \: \dfrac{\mathrm{d}n^+}{\mathrm{d}\mu} \: \dfrac{v}{L}~.
\end{equation}
Here, ${\mathrm{d}n^+}/{\mathrm{d}\mu}$ is the density of states in the wire for electrons moving from left to right.

Let us assume that the considered wire can be described with a square-well potential, Then the energies of the electronic levels are:
\begin{equation}
    E = \dfrac{n^2 h^2}{8 mL^2}~,
\end{equation}
where $m$ is the mass of the charge carrier.
Then, the density of states becomes
\begin{equation} \label{densit}
    \dfrac{\mathrm{d}n^+}{\mathrm{d}\mu} = 2 \times \dfrac{1}{2} \times \dfrac{2L}{h} \: v^{-1}~,
\end{equation} 
where the factor of 2 accounts for the spin of the electrons, while the factor of 1/2 accounts for the fact that we are interested only in the states corresponding to electrons propagating  from left to right. 
Inserting Eq.~\eqref{densit} into Eq.~\eqref{I1}, the current is given by:
\begin{equation}
    I = \dfrac{2e}{h} \: (\mu_\mathrm{L} - \mu_\mathrm{R}) = \dfrac{2e^2}{h} V_\mathrm{b}~, 
\end{equation}
giving the conductance Landauer expression for the conductance for a perfectly-transmitting channel:
\begin{equation}
    G = \dfrac{2e^2}{h}~.
\end{equation}
If the transmission through the wire is less than 1, the conductance can be expressed using Eq.~(25).

\section{Generalized theory far away from resonance \label{appA}}
First, we note that  Eq.~(9) can be equivalently written as:\cite{sowa2018beyond}
\begin{equation}\label{KH}
    K_\pm(\epsilon) = \mathrm{Re} \bigg[\sqrt{\dfrac{\pi}{4\lambda k_{\mathrm{B}} T }} \exp\left( \dfrac{(\Gamma - \mathrm{i}\nu_\pm)^2}{4\lambda k_{\mathrm{B}} T}\right)  \mathrm{erfc}\left(\dfrac{\Gamma - \mathrm{i}\nu_\pm}{\sqrt{4\lambda k_{\mathrm{B}} T}}\right)\bigg] ~,
\end{equation}
where $\nu_\pm = \lambda \mp (\epsilon - \varepsilon_0 + T \Delta S^\circ)$, $\Gamma$ is again the lifetime broadening, and $\mathrm{erfc}(x)$ denotes a complementary error function.
Let us consider here the limit of $\Gamma \ll \sqrt{4\lambda k_\mathrm{B}T}$ and define
\begin{align}\label{xdef}
    x :=& \nu_\pm/\sqrt{4\lambda k_\mathrm{B} T}~;\\
    y :=& \Gamma/\sqrt{4\lambda k_\mathrm{B} T}~.
\end{align}
Then, the molecular densities of states can be written as
\begin{equation} \label{appp3}
    K_\pm(\epsilon) = \sqrt{\dfrac{\pi}{4\lambda k_{\mathrm{B}} T }} \: \mathrm{Re} [w(x+\mathrm{i}y)]~,
\end{equation}
where $w(x+\mathrm{i}y)$ is the Faddeeva function [the real part of which (up to a factor) is the Voigt function], 
\begin{equation} \label{fadeeva}
    w(x+\mathrm{i}y) = \exp[(y -\mathrm{i}x )^2] \: \mathrm{erfc}[y-\mathrm{i}x]~.
\end{equation}
We can now take the limit $y \ll 1$ where Eq.~\eqref{fadeeva} can be approximated [by considering the Taylor expansion of $w(x+\mathrm{i}y)$ around $y=0$] as:\cite{abrarov2015accurate}
\begin{equation}\label{abrarov}
   w(x+\mathrm{i}y) \approx  e^{-x^2} (1-2\mathrm{i}xy)[1 + \mathrm{erf}(\mathrm{i}x)] - \dfrac{2y}{\sqrt{\pi}}~,
\end{equation}
where $\mathrm{erf}(x)$ denotes the error function.
Substituting  Eq.~\eqref{abrarov} to Eq.~\eqref{appp3} gives (after some rearrangements):
\begin{multline}\label{approxm}
    K_\pm(\epsilon) \approx \sqrt{\dfrac{\pi}{4\lambda k_{\mathrm{B}} T}} \exp\left( -\dfrac{[\lambda \mp (\epsilon - \varepsilon_0 +T \Delta S^\circ)]^2}{4\lambda k_{\mathrm{B}} T}\right)  \\ + \dfrac{\Gamma}{\lambda k_{\mathrm{B}} T} \bigg[ \dfrac{\lambda \mp (\epsilon - \varepsilon_0 +T \Delta S^\circ)}{\sqrt{4\lambda k_{\mathrm{B}} T}} \\  D\left(\dfrac{\lambda \mp (\epsilon - \varepsilon_0 +T \Delta S^\circ)}{\sqrt{4\lambda k_{\mathrm{B}} T}}\right) - \dfrac{1}{2} \bigg]~,
\end{multline}
where $D(x)$ denotes a Dawson function.
We note here that the first term on the right-hand side of Eq.~\eqref{approxm} is simply the Marcus rate from Eq.~(27), so that the correction to the conventional Marcus rate is proportional to $\Gamma$.  
Since Eq.~\eqref{KH} was derived for the case of weak molecule-lead coupling,\cite{sowa2018beyond} the (mathematical) validity of Eq.~\eqref{approxm} typically  coincides with the (physical) validity of the generalized Marcus theory.\\
Next, we consider Eq.~\eqref{approxm} in the limit of large $\lvert \epsilon - \varepsilon_0 \rvert$,
\begin{equation}
    \lvert \epsilon - \varepsilon_0 \rvert \gg \lambda, k_\mathrm{B}T, \Gamma, T \Delta S^\circ~.
\end{equation}
Firstly, we note that the first term on the right-hand side of Eq.~\eqref{approxm} vanishes. Secondly, using the definition of $x$ from Eq.~\eqref{xdef}, we note that for large $x$, the Dawson function can be approximated as:
\begin{equation}
    D(x) \approx \dfrac{1}{2x} + \dfrac{1}{4x^3}~.
\end{equation}
Then, 
\begin{equation}
    K_\pm(\epsilon) \approx \dfrac{\Gamma}{\lambda k_{\mathrm{B}} T} \bigg[x\left(\dfrac{1}{2x} + \dfrac{1}{4x^3} \right) - \dfrac{1}{2} \bigg] ~,
\end{equation}
which inserting the definition of $x$ becomes,
\begin{equation}
    K_\pm(\epsilon) \approx \dfrac{\Gamma}{\nu_\pm^2}~.
\end{equation}
Finally, setting
\begin{equation}
    \nu_\pm = [\lambda \mp (\epsilon - \varepsilon_0 + T \Delta S^\circ)]^2 \approx (\epsilon - \varepsilon_0)^2 ~,
\end{equation}
gives Eq.~(28) in the main body of this work.


\begin{thebibliography}{88}%
\makeatletter
\providecommand \@ifxundefined [1]{%
 \@ifx{#1\undefined}
}%
\providecommand \@ifnum [1]{%
 \ifnum #1\expandafter \@firstoftwo
 \else \expandafter \@secondoftwo
 \fi
}%
\providecommand \@ifx [1]{%
 \ifx #1\expandafter \@firstoftwo
 \else \expandafter \@secondoftwo
 \fi
}%
\providecommand \natexlab [1]{#1}%
\providecommand \enquote  [1]{``#1''}%
\providecommand \bibnamefont  [1]{#1}%
\providecommand \bibfnamefont [1]{#1}%
\providecommand \citenamefont [1]{#1}%
\providecommand \href@noop [0]{\@secondoftwo}%
\providecommand \href [0]{\begingroup \@sanitize@url \@href}%
\providecommand \@href[1]{\@@startlink{#1}\@@href}%
\providecommand \@@href[1]{\endgroup#1\@@endlink}%
\providecommand \@sanitize@url [0]{\catcode `\\12\catcode `\$12\catcode
  `\&12\catcode `\#12\catcode `\^12\catcode `\_12\catcode `\%12\relax}%
\providecommand \@@startlink[1]{}%
\providecommand \@@endlink[0]{}%
\providecommand \url  [0]{\begingroup\@sanitize@url \@url }%
\providecommand \@url [1]{\endgroup\@href {#1}{\urlprefix }}%
\providecommand \urlprefix  [0]{URL }%
\providecommand \Eprint [0]{\href }%
\providecommand \doibase [0]{http://dx.doi.org/}%
\providecommand \selectlanguage [0]{\@gobble}%
\providecommand \bibinfo  [0]{\@secondoftwo}%
\providecommand \bibfield  [0]{\@secondoftwo}%
\providecommand \translation [1]{[#1]}%
\providecommand \BibitemOpen [0]{}%
\providecommand \bibitemStop [0]{}%
\providecommand \bibitemNoStop [0]{.\EOS\space}%
\providecommand \EOS [0]{\spacefactor3000\relax}%
\providecommand \BibitemShut  [1]{\csname bibitem#1\endcsname}%
\let\auto@bib@innerbib\@empty
\bibitem [{\citenamefont {Reed}\ \emph {et~al.}(1997)\citenamefont {Reed},
  \citenamefont {Zhou}, \citenamefont {Muller}, \citenamefont {Burgin},\ and\
  \citenamefont {Tour}}]{reed1997conductance}%
  \BibitemOpen
  \bibfield  {author} {\bibinfo {author} {\bibfnamefont {M.~A.}\ \bibnamefont
  {Reed}}, \bibinfo {author} {\bibfnamefont {C.}~\bibnamefont {Zhou}}, \bibinfo
  {author} {\bibfnamefont {C.}~\bibnamefont {Muller}}, \bibinfo {author}
  {\bibfnamefont {T.}~\bibnamefont {Burgin}}, \ and\ \bibinfo {author}
  {\bibfnamefont {J.}~\bibnamefont {Tour}},\ }\href@noop {} {\bibfield
  {journal} {\bibinfo  {journal} {Science}\ }\textbf {\bibinfo {volume}
  {278}},\ \bibinfo {pages} {252} (\bibinfo {year} {1997})}\BibitemShut
  {NoStop}%
\bibitem [{\citenamefont {Park}\ \emph {et~al.}(2000)\citenamefont {Park},
  \citenamefont {Park}, \citenamefont {Lim}, \citenamefont {Anderson},
  \citenamefont {Alivisatos},\ and\ \citenamefont
  {McEuen}}]{park2000nanomechanical}%
  \BibitemOpen
  \bibfield  {author} {\bibinfo {author} {\bibfnamefont {H.}~\bibnamefont
  {Park}}, \bibinfo {author} {\bibfnamefont {J.}~\bibnamefont {Park}}, \bibinfo
  {author} {\bibfnamefont {A.~K.}\ \bibnamefont {Lim}}, \bibinfo {author}
  {\bibfnamefont {E.~H.}\ \bibnamefont {Anderson}}, \bibinfo {author}
  {\bibfnamefont {A.~P.}\ \bibnamefont {Alivisatos}}, \ and\ \bibinfo {author}
  {\bibfnamefont {P.~L.}\ \bibnamefont {McEuen}},\ }\href@noop {} {\bibfield
  {journal} {\bibinfo  {journal} {Nature}\ }\textbf {\bibinfo {volume} {407}},\
  \bibinfo {pages} {57} (\bibinfo {year} {2000})}\BibitemShut {NoStop}%
\bibitem [{\citenamefont {Park}\ \emph {et~al.}(2002)\citenamefont {Park},
  \citenamefont {Pasupathy}, \citenamefont {Goldsmith}, \citenamefont {Chang},
  \citenamefont {Yaish}, \citenamefont {Petta}, \citenamefont {Rinkoski},
  \citenamefont {Sethna}, \citenamefont {Abru{\~n}a}, \citenamefont {McEuen}
  \emph {et~al.}}]{park2002coulomb}%
  \BibitemOpen
  \bibfield  {author} {\bibinfo {author} {\bibfnamefont {J.}~\bibnamefont
  {Park}}, \bibinfo {author} {\bibfnamefont {A.~N.}\ \bibnamefont {Pasupathy}},
  \bibinfo {author} {\bibfnamefont {J.~I.}\ \bibnamefont {Goldsmith}}, \bibinfo
  {author} {\bibfnamefont {C.}~\bibnamefont {Chang}}, \bibinfo {author}
  {\bibfnamefont {Y.}~\bibnamefont {Yaish}}, \bibinfo {author} {\bibfnamefont
  {J.~R.}\ \bibnamefont {Petta}}, \bibinfo {author} {\bibfnamefont
  {M.}~\bibnamefont {Rinkoski}}, \bibinfo {author} {\bibfnamefont {J.~P.}\
  \bibnamefont {Sethna}}, \bibinfo {author} {\bibfnamefont {H.~D.}\
  \bibnamefont {Abru{\~n}a}}, \bibinfo {author} {\bibfnamefont {P.~L.}\
  \bibnamefont {McEuen}},  \emph {et~al.},\ }\href@noop {} {\bibfield
  {journal} {\bibinfo  {journal} {Nature}\ }\textbf {\bibinfo {volume} {417}},\
  \bibinfo {pages} {722} (\bibinfo {year} {2002})}\BibitemShut {NoStop}%
\bibitem [{\citenamefont {Liang}\ \emph {et~al.}(2002)\citenamefont {Liang},
  \citenamefont {Shores}, \citenamefont {Bockrath}, \citenamefont {Long},\ and\
  \citenamefont {Park}}]{liang2002kondo}%
  \BibitemOpen
  \bibfield  {author} {\bibinfo {author} {\bibfnamefont {W.}~\bibnamefont
  {Liang}}, \bibinfo {author} {\bibfnamefont {M.~P.}\ \bibnamefont {Shores}},
  \bibinfo {author} {\bibfnamefont {M.}~\bibnamefont {Bockrath}}, \bibinfo
  {author} {\bibfnamefont {J.~R.}\ \bibnamefont {Long}}, \ and\ \bibinfo
  {author} {\bibfnamefont {H.}~\bibnamefont {Park}},\ }\href@noop {} {\bibfield
   {journal} {\bibinfo  {journal} {Nature}\ }\textbf {\bibinfo {volume}
  {417}},\ \bibinfo {pages} {725} (\bibinfo {year} {2002})}\BibitemShut
  {NoStop}%
\bibitem [{\citenamefont {Cui}\ \emph {et~al.}(2001)\citenamefont {Cui},
  \citenamefont {Primak}, \citenamefont {Zarate}, \citenamefont {Tomfohr},
  \citenamefont {Sankey}, \citenamefont {Moore}, \citenamefont {Moore},
  \citenamefont {Gust}, \citenamefont {Harris},\ and\ \citenamefont
  {Lindsay}}]{cui2001reproducible}%
  \BibitemOpen
  \bibfield  {author} {\bibinfo {author} {\bibfnamefont {X.}~\bibnamefont
  {Cui}}, \bibinfo {author} {\bibfnamefont {A.}~\bibnamefont {Primak}},
  \bibinfo {author} {\bibfnamefont {X.}~\bibnamefont {Zarate}}, \bibinfo
  {author} {\bibfnamefont {J.}~\bibnamefont {Tomfohr}}, \bibinfo {author}
  {\bibfnamefont {O.}~\bibnamefont {Sankey}}, \bibinfo {author} {\bibfnamefont
  {A.}~\bibnamefont {Moore}}, \bibinfo {author} {\bibfnamefont
  {T.}~\bibnamefont {Moore}}, \bibinfo {author} {\bibfnamefont
  {D.}~\bibnamefont {Gust}}, \bibinfo {author} {\bibfnamefont {G.}~\bibnamefont
  {Harris}}, \ and\ \bibinfo {author} {\bibfnamefont {S.}~\bibnamefont
  {Lindsay}},\ }\href@noop {} {\bibfield  {journal} {\bibinfo  {journal}
  {Science}\ }\textbf {\bibinfo {volume} {294}},\ \bibinfo {pages} {571}
  (\bibinfo {year} {2001})}\BibitemShut {NoStop}%
\bibitem [{\citenamefont {Kergueris}\ \emph {et~al.}(1999)\citenamefont
  {Kergueris}, \citenamefont {Bourgoin}, \citenamefont {Palacin}, \citenamefont
  {Est{\`e}ve}, \citenamefont {Urbina}, \citenamefont {Magoga},\ and\
  \citenamefont {Joachim}}]{kergueris1999electron}%
  \BibitemOpen
  \bibfield  {author} {\bibinfo {author} {\bibfnamefont {C.}~\bibnamefont
  {Kergueris}}, \bibinfo {author} {\bibfnamefont {J.-P.}\ \bibnamefont
  {Bourgoin}}, \bibinfo {author} {\bibfnamefont {S.}~\bibnamefont {Palacin}},
  \bibinfo {author} {\bibfnamefont {D.}~\bibnamefont {Est{\`e}ve}}, \bibinfo
  {author} {\bibfnamefont {C.}~\bibnamefont {Urbina}}, \bibinfo {author}
  {\bibfnamefont {M.}~\bibnamefont {Magoga}}, \ and\ \bibinfo {author}
  {\bibfnamefont {C.}~\bibnamefont {Joachim}},\ }\href@noop {} {\bibfield
  {journal} {\bibinfo  {journal} {Phys. Rev. B}\ }\textbf {\bibinfo {volume}
  {59}},\ \bibinfo {pages} {12505} (\bibinfo {year} {1999})}\BibitemShut
  {NoStop}%
\bibitem [{\citenamefont {Perrin}, \citenamefont {Burzur{\'\i}},\ and\
  \citenamefont {van~der Zant}(2015)}]{perrin2015single}%
  \BibitemOpen
  \bibfield  {author} {\bibinfo {author} {\bibfnamefont {M.~L.}\ \bibnamefont
  {Perrin}}, \bibinfo {author} {\bibfnamefont {E.}~\bibnamefont
  {Burzur{\'\i}}}, \ and\ \bibinfo {author} {\bibfnamefont {H.~S.}\
  \bibnamefont {van~der Zant}},\ }\href@noop {} {\bibfield  {journal} {\bibinfo
   {journal} {Chem. Soc. Rev.}\ }\textbf {\bibinfo {volume} {44}},\ \bibinfo
  {pages} {902} (\bibinfo {year} {2015})}\BibitemShut {NoStop}%
\bibitem [{\citenamefont {Gehring}\ \emph {et~al.}(2017)\citenamefont
  {Gehring}, \citenamefont {Sowa}, \citenamefont {Cremers}, \citenamefont {Wu},
  \citenamefont {Sadeghi}, \citenamefont {Sheng}, \citenamefont {Warner},
  \citenamefont {Lambert}, \citenamefont {Briggs},\ and\ \citenamefont
  {Mol}}]{gehring2017distinguishing}%
  \BibitemOpen
  \bibfield  {author} {\bibinfo {author} {\bibfnamefont {P.}~\bibnamefont
  {Gehring}}, \bibinfo {author} {\bibfnamefont {J.~K.}\ \bibnamefont {Sowa}},
  \bibinfo {author} {\bibfnamefont {J.}~\bibnamefont {Cremers}}, \bibinfo
  {author} {\bibfnamefont {Q.}~\bibnamefont {Wu}}, \bibinfo {author}
  {\bibfnamefont {H.}~\bibnamefont {Sadeghi}}, \bibinfo {author} {\bibfnamefont
  {Y.}~\bibnamefont {Sheng}}, \bibinfo {author} {\bibfnamefont {J.~H.}\
  \bibnamefont {Warner}}, \bibinfo {author} {\bibfnamefont {C.~J.}\
  \bibnamefont {Lambert}}, \bibinfo {author} {\bibfnamefont {G.~A.~D.}\
  \bibnamefont {Briggs}}, \ and\ \bibinfo {author} {\bibfnamefont {J.~A.}\
  \bibnamefont {Mol}},\ }\href@noop {} {\bibfield  {journal} {\bibinfo
  {journal} {ACS Nano}\ }\textbf {\bibinfo {volume} {11}},\ \bibinfo {pages}
  {5325} (\bibinfo {year} {2017})}\BibitemShut {NoStop}%
\bibitem [{\citenamefont {Elbing}\ \emph {et~al.}(2005)\citenamefont {Elbing},
  \citenamefont {Ochs}, \citenamefont {Koentopp}, \citenamefont {Fischer},
  \citenamefont {von H{\"a}nisch}, \citenamefont {Weigend}, \citenamefont
  {Evers}, \citenamefont {Weber},\ and\ \citenamefont
  {Mayor}}]{elbing2005single}%
  \BibitemOpen
  \bibfield  {author} {\bibinfo {author} {\bibfnamefont {M.}~\bibnamefont
  {Elbing}}, \bibinfo {author} {\bibfnamefont {R.}~\bibnamefont {Ochs}},
  \bibinfo {author} {\bibfnamefont {M.}~\bibnamefont {Koentopp}}, \bibinfo
  {author} {\bibfnamefont {M.}~\bibnamefont {Fischer}}, \bibinfo {author}
  {\bibfnamefont {C.}~\bibnamefont {von H{\"a}nisch}}, \bibinfo {author}
  {\bibfnamefont {F.}~\bibnamefont {Weigend}}, \bibinfo {author} {\bibfnamefont
  {F.}~\bibnamefont {Evers}}, \bibinfo {author} {\bibfnamefont {H.~B.}\
  \bibnamefont {Weber}}, \ and\ \bibinfo {author} {\bibfnamefont
  {M.}~\bibnamefont {Mayor}},\ }\href@noop {} {\bibfield  {journal} {\bibinfo
  {journal} {Proc. Natl. Acad. Sci. U.S.A.}\ }\textbf {\bibinfo {volume}
  {102}},\ \bibinfo {pages} {8815} (\bibinfo {year} {2005})}\BibitemShut
  {NoStop}%
\bibitem [{\citenamefont {D{\'\i}ez-P{\'e}rez}\ \emph
  {et~al.}(2009)\citenamefont {D{\'\i}ez-P{\'e}rez}, \citenamefont {Hihath},
  \citenamefont {Lee}, \citenamefont {Yu}, \citenamefont {Adamska},
  \citenamefont {Kozhushner}, \citenamefont {Oleynik},\ and\ \citenamefont
  {Tao}}]{diez2009rectification}%
  \BibitemOpen
  \bibfield  {author} {\bibinfo {author} {\bibfnamefont {I.}~\bibnamefont
  {D{\'\i}ez-P{\'e}rez}}, \bibinfo {author} {\bibfnamefont {J.}~\bibnamefont
  {Hihath}}, \bibinfo {author} {\bibfnamefont {Y.}~\bibnamefont {Lee}},
  \bibinfo {author} {\bibfnamefont {L.}~\bibnamefont {Yu}}, \bibinfo {author}
  {\bibfnamefont {L.}~\bibnamefont {Adamska}}, \bibinfo {author} {\bibfnamefont
  {M.~A.}\ \bibnamefont {Kozhushner}}, \bibinfo {author} {\bibfnamefont
  {I.~I.}\ \bibnamefont {Oleynik}}, \ and\ \bibinfo {author} {\bibfnamefont
  {N.}~\bibnamefont {Tao}},\ }\href@noop {} {\bibfield  {journal} {\bibinfo
  {journal} {Nat. Chem.}\ }\textbf {\bibinfo {volume} {1}},\ \bibinfo {pages}
  {635} (\bibinfo {year} {2009})}\BibitemShut {NoStop}%
\bibitem [{\citenamefont {Perrin}\ \emph {et~al.}(2016)\citenamefont {Perrin},
  \citenamefont {Gal{\'a}n}, \citenamefont {Eelkema}, \citenamefont {Thijssen},
  \citenamefont {Grozema},\ and\ \citenamefont {van~der
  Zant}}]{perrin2016gate}%
  \BibitemOpen
  \bibfield  {author} {\bibinfo {author} {\bibfnamefont {M.~L.}\ \bibnamefont
  {Perrin}}, \bibinfo {author} {\bibfnamefont {E.}~\bibnamefont {Gal{\'a}n}},
  \bibinfo {author} {\bibfnamefont {R.}~\bibnamefont {Eelkema}}, \bibinfo
  {author} {\bibfnamefont {J.~M.}\ \bibnamefont {Thijssen}}, \bibinfo {author}
  {\bibfnamefont {F.}~\bibnamefont {Grozema}}, \ and\ \bibinfo {author}
  {\bibfnamefont {H.~S.}\ \bibnamefont {van~der Zant}},\ }\href@noop {}
  {\bibfield  {journal} {\bibinfo  {journal} {Nanoscale}\ }\textbf {\bibinfo
  {volume} {8}},\ \bibinfo {pages} {8919} (\bibinfo {year} {2016})}\BibitemShut
  {NoStop}%
\bibitem [{\citenamefont {Iacovita}\ \emph {et~al.}(2008)\citenamefont
  {Iacovita}, \citenamefont {Rastei}, \citenamefont {Heinrich}, \citenamefont
  {Brumme}, \citenamefont {Kortus}, \citenamefont {Limot},\ and\ \citenamefont
  {Bucher}}]{iacovita2008visualizing}%
  \BibitemOpen
  \bibfield  {author} {\bibinfo {author} {\bibfnamefont {C.}~\bibnamefont
  {Iacovita}}, \bibinfo {author} {\bibfnamefont {M.}~\bibnamefont {Rastei}},
  \bibinfo {author} {\bibfnamefont {B.}~\bibnamefont {Heinrich}}, \bibinfo
  {author} {\bibfnamefont {T.}~\bibnamefont {Brumme}}, \bibinfo {author}
  {\bibfnamefont {J.}~\bibnamefont {Kortus}}, \bibinfo {author} {\bibfnamefont
  {L.}~\bibnamefont {Limot}}, \ and\ \bibinfo {author} {\bibfnamefont
  {J.}~\bibnamefont {Bucher}},\ }\href@noop {} {\bibfield  {journal} {\bibinfo
  {journal} {Phys. Rev. Lett.}\ }\textbf {\bibinfo {volume} {101}},\ \bibinfo
  {pages} {116602} (\bibinfo {year} {2008})}\BibitemShut {NoStop}%
\bibitem [{\citenamefont {Sanvito}(2011)}]{sanvito2011molecular}%
  \BibitemOpen
  \bibfield  {author} {\bibinfo {author} {\bibfnamefont {S.}~\bibnamefont
  {Sanvito}},\ }\href@noop {} {\bibfield  {journal} {\bibinfo  {journal} {Chem.
  Soc. Rev.}\ }\textbf {\bibinfo {volume} {40}},\ \bibinfo {pages} {3336}
  (\bibinfo {year} {2011})}\BibitemShut {NoStop}%
\bibitem [{\citenamefont {Bogani}\ and\ \citenamefont
  {Wernsdorfer}(2008)}]{bogani2008molecular}%
  \BibitemOpen
  \bibfield  {author} {\bibinfo {author} {\bibfnamefont {L.}~\bibnamefont
  {Bogani}}\ and\ \bibinfo {author} {\bibfnamefont {W.}~\bibnamefont
  {Wernsdorfer}},\ }\href@noop {} {\bibfield  {journal} {\bibinfo  {journal}
  {Nat. Mater.}\ }\textbf {\bibinfo {volume} {7}},\ \bibinfo {pages} {179}
  (\bibinfo {year} {2008})}\BibitemShut {NoStop}%
\bibitem [{\citenamefont {Reddy}\ \emph {et~al.}(2007)\citenamefont {Reddy},
  \citenamefont {Jang}, \citenamefont {Segalman},\ and\ \citenamefont
  {Majumdar}}]{reddy2007thermoelectricity}%
  \BibitemOpen
  \bibfield  {author} {\bibinfo {author} {\bibfnamefont {P.}~\bibnamefont
  {Reddy}}, \bibinfo {author} {\bibfnamefont {S.-Y.}\ \bibnamefont {Jang}},
  \bibinfo {author} {\bibfnamefont {R.~A.}\ \bibnamefont {Segalman}}, \ and\
  \bibinfo {author} {\bibfnamefont {A.}~\bibnamefont {Majumdar}},\ }\href@noop
  {} {\bibfield  {journal} {\bibinfo  {journal} {Science}\ }\textbf {\bibinfo
  {volume} {315}},\ \bibinfo {pages} {1568} (\bibinfo {year}
  {2007})}\BibitemShut {NoStop}%
\bibitem [{\citenamefont {Cui}\ \emph {et~al.}(2017)\citenamefont {Cui},
  \citenamefont {Miao}, \citenamefont {Jiang}, \citenamefont {Meyhofer},\ and\
  \citenamefont {Reddy}}]{cui2017perspective}%
  \BibitemOpen
  \bibfield  {author} {\bibinfo {author} {\bibfnamefont {L.}~\bibnamefont
  {Cui}}, \bibinfo {author} {\bibfnamefont {R.}~\bibnamefont {Miao}}, \bibinfo
  {author} {\bibfnamefont {C.}~\bibnamefont {Jiang}}, \bibinfo {author}
  {\bibfnamefont {E.}~\bibnamefont {Meyhofer}}, \ and\ \bibinfo {author}
  {\bibfnamefont {P.}~\bibnamefont {Reddy}},\ }\href@noop {} {\bibfield
  {journal} {\bibinfo  {journal} {J. Chem. Phys.}\ }\textbf {\bibinfo {volume}
  {146}},\ \bibinfo {pages} {092201} (\bibinfo {year} {2017})}\BibitemShut
  {NoStop}%
\bibitem [{\citenamefont {Chen}\ \emph {et~al.}(2017)\citenamefont {Chen},
  \citenamefont {Roemer}, \citenamefont {Yuan}, \citenamefont {Du},
  \citenamefont {Thompson}, \citenamefont {del Barco},\ and\ \citenamefont
  {Nijhuis}}]{chen2017molecular}%
  \BibitemOpen
  \bibfield  {author} {\bibinfo {author} {\bibfnamefont {X.}~\bibnamefont
  {Chen}}, \bibinfo {author} {\bibfnamefont {M.}~\bibnamefont {Roemer}},
  \bibinfo {author} {\bibfnamefont {L.}~\bibnamefont {Yuan}}, \bibinfo {author}
  {\bibfnamefont {W.}~\bibnamefont {Du}}, \bibinfo {author} {\bibfnamefont
  {D.}~\bibnamefont {Thompson}}, \bibinfo {author} {\bibfnamefont
  {E.}~\bibnamefont {del Barco}}, \ and\ \bibinfo {author} {\bibfnamefont
  {C.~A.}\ \bibnamefont {Nijhuis}},\ }\href@noop {} {\bibfield  {journal}
  {\bibinfo  {journal} {Nat. Nanotechnol.}\ }\textbf {\bibinfo {volume} {12}},\
  \bibinfo {pages} {797} (\bibinfo {year} {2017})}\BibitemShut {NoStop}%
\bibitem [{\citenamefont {Nitzan}(2001)}]{nitzan2001electron}%
  \BibitemOpen
  \bibfield  {author} {\bibinfo {author} {\bibfnamefont {A.}~\bibnamefont
  {Nitzan}},\ }\href@noop {} {\bibfield  {journal} {\bibinfo  {journal} {Annu.
  Rev. Phys. Chem.}\ }\textbf {\bibinfo {volume} {52}},\ \bibinfo {pages} {681}
  (\bibinfo {year} {2001})}\BibitemShut {NoStop}%
\bibitem [{\citenamefont {Engelkes}, \citenamefont {Beebe},\ and\ \citenamefont
  {Frisbie}(2004)}]{engelkes2004length}%
  \BibitemOpen
  \bibfield  {author} {\bibinfo {author} {\bibfnamefont {V.~B.}\ \bibnamefont
  {Engelkes}}, \bibinfo {author} {\bibfnamefont {J.~M.}\ \bibnamefont {Beebe}},
  \ and\ \bibinfo {author} {\bibfnamefont {C.~D.}\ \bibnamefont {Frisbie}},\
  }\href@noop {} {\bibfield  {journal} {\bibinfo  {journal} {J. Am. Chem.
  Soc.}\ }\textbf {\bibinfo {volume} {126}},\ \bibinfo {pages} {14287}
  (\bibinfo {year} {2004})}\BibitemShut {NoStop}%
\bibitem [{\citenamefont {Lindsay}\ and\ \citenamefont
  {Ratner}(2007)}]{lindsay2007molecular}%
  \BibitemOpen
  \bibfield  {author} {\bibinfo {author} {\bibfnamefont {S.~M.}\ \bibnamefont
  {Lindsay}}\ and\ \bibinfo {author} {\bibfnamefont {M.~A.}\ \bibnamefont
  {Ratner}},\ }\href@noop {} {\bibfield  {journal} {\bibinfo  {journal} {Adv.
  Mater.}\ }\textbf {\bibinfo {volume} {19}},\ \bibinfo {pages} {23} (\bibinfo
  {year} {2007})}\BibitemShut {NoStop}%
\bibitem [{\citenamefont {Jin}\ \emph {et~al.}(2013)\citenamefont {Jin},
  \citenamefont {Strange}, \citenamefont {Markussen}, \citenamefont {Solomon},\
  and\ \citenamefont {Thygesen}}]{jin2013energy}%
  \BibitemOpen
  \bibfield  {author} {\bibinfo {author} {\bibfnamefont {C.}~\bibnamefont
  {Jin}}, \bibinfo {author} {\bibfnamefont {M.}~\bibnamefont {Strange}},
  \bibinfo {author} {\bibfnamefont {T.}~\bibnamefont {Markussen}}, \bibinfo
  {author} {\bibfnamefont {G.~C.}\ \bibnamefont {Solomon}}, \ and\ \bibinfo
  {author} {\bibfnamefont {K.~S.}\ \bibnamefont {Thygesen}},\ }\href@noop {}
  {\bibfield  {journal} {\bibinfo  {journal} {J. Chem. Phys.}\ }\textbf
  {\bibinfo {volume} {139}},\ \bibinfo {pages} {184307} (\bibinfo {year}
  {2013})}\BibitemShut {NoStop}%
\bibitem [{\citenamefont {Secker}\ \emph {et~al.}(2011)\citenamefont {Secker},
  \citenamefont {Wagner}, \citenamefont {Ballmann}, \citenamefont {H{\"a}rtle},
  \citenamefont {Thoss},\ and\ \citenamefont {Weber}}]{secker2011resonant}%
  \BibitemOpen
  \bibfield  {author} {\bibinfo {author} {\bibfnamefont {D.}~\bibnamefont
  {Secker}}, \bibinfo {author} {\bibfnamefont {S.}~\bibnamefont {Wagner}},
  \bibinfo {author} {\bibfnamefont {S.}~\bibnamefont {Ballmann}}, \bibinfo
  {author} {\bibfnamefont {R.}~\bibnamefont {H{\"a}rtle}}, \bibinfo {author}
  {\bibfnamefont {M.}~\bibnamefont {Thoss}}, \ and\ \bibinfo {author}
  {\bibfnamefont {H.~B.}\ \bibnamefont {Weber}},\ }\href@noop {} {\bibfield
  {journal} {\bibinfo  {journal} {Phys. Rev. Lett.}\ }\textbf {\bibinfo
  {volume} {106}},\ \bibinfo {pages} {136807} (\bibinfo {year}
  {2011})}\BibitemShut {NoStop}%
\bibitem [{\citenamefont {Burzur{\'\i}}\ \emph {et~al.}(2016)\citenamefont
  {Burzur{\'\i}}, \citenamefont {Island}, \citenamefont {D{\'\i}az-Torres},
  \citenamefont {Fursina}, \citenamefont {Gonz\'{a}lez-Campo}, \citenamefont
  {Roubeau}, \citenamefont {Teat}, \citenamefont {Aliaga-Alcalde},
  \citenamefont {Ruiz},\ and\ \citenamefont {van~der
  Zant}}]{burzuri2016sequential}%
  \BibitemOpen
  \bibfield  {author} {\bibinfo {author} {\bibfnamefont {E.}~\bibnamefont
  {Burzur{\'\i}}}, \bibinfo {author} {\bibfnamefont {J.~O.}\ \bibnamefont
  {Island}}, \bibinfo {author} {\bibfnamefont {R.}~\bibnamefont
  {D{\'\i}az-Torres}}, \bibinfo {author} {\bibfnamefont {A.}~\bibnamefont
  {Fursina}}, \bibinfo {author} {\bibfnamefont {A.}~\bibnamefont
  {Gonz\'{a}lez-Campo}}, \bibinfo {author} {\bibfnamefont {O.}~\bibnamefont
  {Roubeau}}, \bibinfo {author} {\bibfnamefont {S.~J.}\ \bibnamefont {Teat}},
  \bibinfo {author} {\bibfnamefont {N.}~\bibnamefont {Aliaga-Alcalde}},
  \bibinfo {author} {\bibfnamefont {E.}~\bibnamefont {Ruiz}}, \ and\ \bibinfo
  {author} {\bibfnamefont {H.~S.~J.}\ \bibnamefont {van~der Zant}},\
  }\href@noop {} {\bibfield  {journal} {\bibinfo  {journal} {ACS Nano}\
  }\textbf {\bibinfo {volume} {10}},\ \bibinfo {pages} {2521} (\bibinfo {year}
  {2016})}\BibitemShut {NoStop}%
\bibitem [{\citenamefont {Thomas}\ \emph {et~al.}(2019)\citenamefont {Thomas},
  \citenamefont {Limburg}, \citenamefont {Sowa}, \citenamefont {Willick},
  \citenamefont {Baugh}, \citenamefont {Briggs}, \citenamefont {Gauger},
  \citenamefont {Anderson},\ and\ \citenamefont
  {Mol}}]{thomas2019understanding}%
  \BibitemOpen
  \bibfield  {author} {\bibinfo {author} {\bibfnamefont {J.~O.}\ \bibnamefont
  {Thomas}}, \bibinfo {author} {\bibfnamefont {B.}~\bibnamefont {Limburg}},
  \bibinfo {author} {\bibfnamefont {J.~K.}\ \bibnamefont {Sowa}}, \bibinfo
  {author} {\bibfnamefont {K.}~\bibnamefont {Willick}}, \bibinfo {author}
  {\bibfnamefont {J.}~\bibnamefont {Baugh}}, \bibinfo {author} {\bibfnamefont
  {G.~A.~D.}\ \bibnamefont {Briggs}}, \bibinfo {author} {\bibfnamefont {E.~M.}\
  \bibnamefont {Gauger}}, \bibinfo {author} {\bibfnamefont {H.~L.}\
  \bibnamefont {Anderson}}, \ and\ \bibinfo {author} {\bibfnamefont {J.~A.}\
  \bibnamefont {Mol}},\ }\href@noop {} {\bibfield  {journal} {\bibinfo
  {journal} {Nat. Commun.}\ }\textbf {\bibinfo {volume} {10}},\ \bibinfo
  {pages} {1} (\bibinfo {year} {2019})}\BibitemShut {NoStop}%
\bibitem [{\citenamefont {Fung}\ \emph {et~al.}(2019)\citenamefont {Fung},
  \citenamefont {Gelbwaser}, \citenamefont {Taylor}, \citenamefont {Low},
  \citenamefont {Xia}, \citenamefont {Davydenko}, \citenamefont {Campos},
  \citenamefont {Marder}, \citenamefont {Peskin},\ and\ \citenamefont
  {Venkataraman}}]{fung2019breaking}%
  \BibitemOpen
  \bibfield  {author} {\bibinfo {author} {\bibfnamefont {E.-D.}\ \bibnamefont
  {Fung}}, \bibinfo {author} {\bibfnamefont {D.}~\bibnamefont {Gelbwaser}},
  \bibinfo {author} {\bibfnamefont {J.}~\bibnamefont {Taylor}}, \bibinfo
  {author} {\bibfnamefont {J.}~\bibnamefont {Low}}, \bibinfo {author}
  {\bibfnamefont {J.}~\bibnamefont {Xia}}, \bibinfo {author} {\bibfnamefont
  {I.}~\bibnamefont {Davydenko}}, \bibinfo {author} {\bibfnamefont {L.~M.}\
  \bibnamefont {Campos}}, \bibinfo {author} {\bibfnamefont {S.}~\bibnamefont
  {Marder}}, \bibinfo {author} {\bibfnamefont {U.}~\bibnamefont {Peskin}}, \
  and\ \bibinfo {author} {\bibfnamefont {L.}~\bibnamefont {Venkataraman}},\
  }\href@noop {} {\bibfield  {journal} {\bibinfo  {journal} {Nano Lett.}\
  }\textbf {\bibinfo {volume} {19}},\ \bibinfo {pages} {2555} (\bibinfo {year}
  {2019})}\BibitemShut {NoStop}%
\bibitem [{\citenamefont {Friis}\ \emph {et~al.}(1998)\citenamefont {Friis},
  \citenamefont {Kharkats}, \citenamefont {Kuznetsov},\ and\ \citenamefont
  {Ulstrup}}]{friis1998situ}%
  \BibitemOpen
  \bibfield  {author} {\bibinfo {author} {\bibfnamefont {E.~P.}\ \bibnamefont
  {Friis}}, \bibinfo {author} {\bibfnamefont {Y.~I.}\ \bibnamefont {Kharkats}},
  \bibinfo {author} {\bibfnamefont {A.~M.}\ \bibnamefont {Kuznetsov}}, \ and\
  \bibinfo {author} {\bibfnamefont {J.}~\bibnamefont {Ulstrup}},\ }\href@noop
  {} {\bibfield  {journal} {\bibinfo  {journal} {J. Phys. Chem. A}\ }\textbf
  {\bibinfo {volume} {102}},\ \bibinfo {pages} {7851} (\bibinfo {year}
  {1998})}\BibitemShut {NoStop}%
\bibitem [{\citenamefont {Kuznetsov}\ and\ \citenamefont
  {Ulstrup}(2000)}]{kuznetsov2000mechanisms}%
  \BibitemOpen
  \bibfield  {author} {\bibinfo {author} {\bibfnamefont {A.~M.}\ \bibnamefont
  {Kuznetsov}}\ and\ \bibinfo {author} {\bibfnamefont {J.}~\bibnamefont
  {Ulstrup}},\ }\href@noop {} {\bibfield  {journal} {\bibinfo  {journal} {J.
  Phys. Chem. A}\ }\textbf {\bibinfo {volume} {104}},\ \bibinfo {pages} {11531}
  (\bibinfo {year} {2000})}\BibitemShut {NoStop}%
\bibitem [{\citenamefont {Kuznetsov}\ and\ \citenamefont
  {Ulstrup}(2002)}]{kuznetsov2002mechanisms}%
  \BibitemOpen
  \bibfield  {author} {\bibinfo {author} {\bibfnamefont {A.~M.}\ \bibnamefont
  {Kuznetsov}}\ and\ \bibinfo {author} {\bibfnamefont {J.}~\bibnamefont
  {Ulstrup}},\ }\href@noop {} {\bibfield  {journal} {\bibinfo  {journal} {J.
  Chem. Phys.}\ }\textbf {\bibinfo {volume} {116}},\ \bibinfo {pages} {2149}
  (\bibinfo {year} {2002})}\BibitemShut {NoStop}%
\bibitem [{\citenamefont {Zhang}\ \emph {et~al.}(2008)\citenamefont {Zhang},
  \citenamefont {Kuznetsov}, \citenamefont {Medvedev}, \citenamefont {Chi},
  \citenamefont {Albrecht}, \citenamefont {Jensen},\ and\ \citenamefont
  {Ulstrup}}]{zhang2008single}%
  \BibitemOpen
  \bibfield  {author} {\bibinfo {author} {\bibfnamefont {J.}~\bibnamefont
  {Zhang}}, \bibinfo {author} {\bibfnamefont {A.~M.}\ \bibnamefont
  {Kuznetsov}}, \bibinfo {author} {\bibfnamefont {I.~G.}\ \bibnamefont
  {Medvedev}}, \bibinfo {author} {\bibfnamefont {Q.}~\bibnamefont {Chi}},
  \bibinfo {author} {\bibfnamefont {T.}~\bibnamefont {Albrecht}}, \bibinfo
  {author} {\bibfnamefont {P.~S.}\ \bibnamefont {Jensen}}, \ and\ \bibinfo
  {author} {\bibfnamefont {J.}~\bibnamefont {Ulstrup}},\ }\href@noop {}
  {\bibfield  {journal} {\bibinfo  {journal} {Chem. Rev.}\ }\textbf {\bibinfo
  {volume} {108}},\ \bibinfo {pages} {2737} (\bibinfo {year}
  {2008})}\BibitemShut {NoStop}%
\bibitem [{\citenamefont {Migliore}\ and\ \citenamefont
  {Nitzan}(2011)}]{migliore2011nonlinear}%
  \BibitemOpen
  \bibfield  {author} {\bibinfo {author} {\bibfnamefont {A.}~\bibnamefont
  {Migliore}}\ and\ \bibinfo {author} {\bibfnamefont {A.}~\bibnamefont
  {Nitzan}},\ }\href@noop {} {\bibfield  {journal} {\bibinfo  {journal} {ACS
  Nano}\ }\textbf {\bibinfo {volume} {5}},\ \bibinfo {pages} {6669} (\bibinfo
  {year} {2011})}\BibitemShut {NoStop}%
\bibitem [{\citenamefont {Migliore}\ and\ \citenamefont
  {Nitzan}(2013)}]{migliore2013irreversibility}%
  \BibitemOpen
  \bibfield  {author} {\bibinfo {author} {\bibfnamefont {A.}~\bibnamefont
  {Migliore}}\ and\ \bibinfo {author} {\bibfnamefont {A.}~\bibnamefont
  {Nitzan}},\ }\href@noop {} {\bibfield  {journal} {\bibinfo  {journal} {J. Am.
  Chem. Soc.}\ }\textbf {\bibinfo {volume} {135}},\ \bibinfo {pages} {9420}
  (\bibinfo {year} {2013})}\BibitemShut {NoStop}%
\bibitem [{\citenamefont {Yuan}\ \emph {et~al.}(2018)\citenamefont {Yuan},
  \citenamefont {Wang}, \citenamefont {Garrigues}, \citenamefont {Jiang},
  \citenamefont {Annadata}, \citenamefont {Antonana}, \citenamefont {Barco},\
  and\ \citenamefont {Nijhuis}}]{yuan2018transition}%
  \BibitemOpen
  \bibfield  {author} {\bibinfo {author} {\bibfnamefont {L.}~\bibnamefont
  {Yuan}}, \bibinfo {author} {\bibfnamefont {L.}~\bibnamefont {Wang}}, \bibinfo
  {author} {\bibfnamefont {A.~R.}\ \bibnamefont {Garrigues}}, \bibinfo {author}
  {\bibfnamefont {L.}~\bibnamefont {Jiang}}, \bibinfo {author} {\bibfnamefont
  {H.~V.}\ \bibnamefont {Annadata}}, \bibinfo {author} {\bibfnamefont {M.~A.}\
  \bibnamefont {Antonana}}, \bibinfo {author} {\bibfnamefont {E.}~\bibnamefont
  {Barco}}, \ and\ \bibinfo {author} {\bibfnamefont {C.~A.}\ \bibnamefont
  {Nijhuis}},\ }\href@noop {} {\bibfield  {journal} {\bibinfo  {journal} {Nat.
  Nanotechnol.}\ }\textbf {\bibinfo {volume} {13}},\ \bibinfo {pages} {322}
  (\bibinfo {year} {2018})}\BibitemShut {NoStop}%
\bibitem [{\citenamefont {Jia}\ \emph {et~al.}(2016)\citenamefont {Jia},
  \citenamefont {Migliore}, \citenamefont {Xin}, \citenamefont {Huang},
  \citenamefont {Wang}, \citenamefont {Yang}, \citenamefont {Wang},
  \citenamefont {Chen}, \citenamefont {Wang}, \citenamefont {Feng} \emph
  {et~al.}}]{jia2016covalently}%
  \BibitemOpen
  \bibfield  {author} {\bibinfo {author} {\bibfnamefont {C.}~\bibnamefont
  {Jia}}, \bibinfo {author} {\bibfnamefont {A.}~\bibnamefont {Migliore}},
  \bibinfo {author} {\bibfnamefont {N.}~\bibnamefont {Xin}}, \bibinfo {author}
  {\bibfnamefont {S.}~\bibnamefont {Huang}}, \bibinfo {author} {\bibfnamefont
  {J.}~\bibnamefont {Wang}}, \bibinfo {author} {\bibfnamefont {Q.}~\bibnamefont
  {Yang}}, \bibinfo {author} {\bibfnamefont {S.}~\bibnamefont {Wang}}, \bibinfo
  {author} {\bibfnamefont {H.}~\bibnamefont {Chen}}, \bibinfo {author}
  {\bibfnamefont {D.}~\bibnamefont {Wang}}, \bibinfo {author} {\bibfnamefont
  {B.}~\bibnamefont {Feng}},  \emph {et~al.},\ }\href@noop {} {\bibfield
  {journal} {\bibinfo  {journal} {Science}\ }\textbf {\bibinfo {volume}
  {352}},\ \bibinfo {pages} {1443} (\bibinfo {year} {2016})}\BibitemShut
  {NoStop}%
\bibitem [{\citenamefont {Sowa}\ \emph {et~al.}(2018)\citenamefont {Sowa},
  \citenamefont {Mol}, \citenamefont {Briggs},\ and\ \citenamefont
  {Gauger}}]{sowa2018beyond}%
  \BibitemOpen
  \bibfield  {author} {\bibinfo {author} {\bibfnamefont {J.~K.}\ \bibnamefont
  {Sowa}}, \bibinfo {author} {\bibfnamefont {J.~A.}\ \bibnamefont {Mol}},
  \bibinfo {author} {\bibfnamefont {G.~A.~D.}\ \bibnamefont {Briggs}}, \ and\
  \bibinfo {author} {\bibfnamefont {E.~M.}\ \bibnamefont {Gauger}},\
  }\href@noop {} {\bibfield  {journal} {\bibinfo  {journal} {J. Chem. Phys.}\
  }\textbf {\bibinfo {volume} {149}},\ \bibinfo {pages} {154112} (\bibinfo
  {year} {2018})}\BibitemShut {NoStop}%
\bibitem [{\citenamefont {Sowa}\ \emph {et~al.}(2020)\citenamefont {Sowa},
  \citenamefont {Lambert}, \citenamefont {Seideman},\ and\ \citenamefont
  {Gauger}}]{sowa2020beyond}%
  \BibitemOpen
  \bibfield  {author} {\bibinfo {author} {\bibfnamefont {J.~K.}\ \bibnamefont
  {Sowa}}, \bibinfo {author} {\bibfnamefont {N.}~\bibnamefont {Lambert}},
  \bibinfo {author} {\bibfnamefont {T.}~\bibnamefont {Seideman}}, \ and\
  \bibinfo {author} {\bibfnamefont {E.~M.}\ \bibnamefont {Gauger}},\
  }\href@noop {} {\bibfield  {journal} {\bibinfo  {journal} {J. Chem. Phys.}\
  }\textbf {\bibinfo {volume} {152}},\ \bibinfo {pages} {064103} (\bibinfo
  {year} {2020})}\BibitemShut {NoStop}%
\bibitem [{\citenamefont {Liu}\ and\ \citenamefont
  {Segal}(2020)}]{liu2020generalized}%
  \BibitemOpen
  \bibfield  {author} {\bibinfo {author} {\bibfnamefont {J.}~\bibnamefont
  {Liu}}\ and\ \bibinfo {author} {\bibfnamefont {D.}~\bibnamefont {Segal}},\
  }\href@noop {} {\bibfield  {journal} {\bibinfo  {journal} {Phys. Rev. B}\
  }\textbf {\bibinfo {volume} {101}},\ \bibinfo {pages} {155407} (\bibinfo
  {year} {2020})}\BibitemShut {NoStop}%
\bibitem [{\citenamefont {Marcus}\ and\ \citenamefont
  {Sutin}(1985)}]{marcus1985electron}%
  \BibitemOpen
  \bibfield  {author} {\bibinfo {author} {\bibfnamefont {R.~A.}\ \bibnamefont
  {Marcus}}\ and\ \bibinfo {author} {\bibfnamefont {N.}~\bibnamefont {Sutin}},\
  }\href@noop {} {\bibfield  {journal} {\bibinfo  {journal} {Biochim. Biophys.
  Acta}\ }\textbf {\bibinfo {volume} {811}},\ \bibinfo {pages} {265} (\bibinfo
  {year} {1985})}\BibitemShut {NoStop}%
\bibitem [{\citenamefont {Barbara}, \citenamefont {Meyer},\ and\ \citenamefont
  {Ratner}(1996)}]{barbara1996contemporary}%
  \BibitemOpen
  \bibfield  {author} {\bibinfo {author} {\bibfnamefont {P.~F.}\ \bibnamefont
  {Barbara}}, \bibinfo {author} {\bibfnamefont {T.~J.}\ \bibnamefont {Meyer}},
  \ and\ \bibinfo {author} {\bibfnamefont {M.~A.}\ \bibnamefont {Ratner}},\
  }\href@noop {} {\bibfield  {journal} {\bibinfo  {journal} {J. Phys. Chem.}\
  }\textbf {\bibinfo {volume} {100}},\ \bibinfo {pages} {13148} (\bibinfo
  {year} {1996})}\BibitemShut {NoStop}%
\bibitem [{\citenamefont {May}\ and\ \citenamefont
  {K\"{u}hn}(2008)}]{may2008charge}%
  \BibitemOpen
  \bibfield  {author} {\bibinfo {author} {\bibfnamefont {V.}~\bibnamefont
  {May}}\ and\ \bibinfo {author} {\bibfnamefont {O.}~\bibnamefont {K\"{u}hn}},\
  }\href@noop {} {\emph {\bibinfo {title} {Charge and energy transfer dynamics
  in molecular systems}}}\ (\bibinfo  {publisher} {John Wiley \& Sons},\
  \bibinfo {year} {2008})\BibitemShut {NoStop}%
\bibitem [{\citenamefont {Kang}\ \emph {et~al.}(2020)\citenamefont {Kang},
  \citenamefont {Kong}, \citenamefont {Byeon}, \citenamefont {Yang},
  \citenamefont {Kim},\ and\ \citenamefont {Yoon}}]{kang2020interplay}%
  \BibitemOpen
  \bibfield  {author} {\bibinfo {author} {\bibfnamefont {H.}~\bibnamefont
  {Kang}}, \bibinfo {author} {\bibfnamefont {G.~D.}\ \bibnamefont {Kong}},
  \bibinfo {author} {\bibfnamefont {S.~E.}\ \bibnamefont {Byeon}}, \bibinfo
  {author} {\bibfnamefont {S.}~\bibnamefont {Yang}}, \bibinfo {author}
  {\bibfnamefont {J.~W.}\ \bibnamefont {Kim}}, \ and\ \bibinfo {author}
  {\bibfnamefont {H.~J.}\ \bibnamefont {Yoon}},\ }\href@noop {} {\bibfield
  {journal} {\bibinfo  {journal} {J. Phys. Chem. Lett.}\ }\textbf {\bibinfo
  {volume} {11}},\ \bibinfo {pages} {8597} (\bibinfo {year}
  {2020})}\BibitemShut {NoStop}%
\bibitem [{\citenamefont {Galperin}, \citenamefont {Nitzan},\ and\
  \citenamefont {Ratner}(2006)}]{galperin2006resonant}%
  \BibitemOpen
  \bibfield  {author} {\bibinfo {author} {\bibfnamefont {M.}~\bibnamefont
  {Galperin}}, \bibinfo {author} {\bibfnamefont {A.}~\bibnamefont {Nitzan}}, \
  and\ \bibinfo {author} {\bibfnamefont {M.~A.}\ \bibnamefont {Ratner}},\
  }\href@noop {} {\bibfield  {journal} {\bibinfo  {journal} {Phys. Rev. B}\
  }\textbf {\bibinfo {volume} {73}},\ \bibinfo {pages} {045314} (\bibinfo
  {year} {2006})}\BibitemShut {NoStop}%
\bibitem [{\citenamefont {Chidsey}(1991)}]{chidsey1991free}%
  \BibitemOpen
  \bibfield  {author} {\bibinfo {author} {\bibfnamefont {C.~E.}\ \bibnamefont
  {Chidsey}},\ }\href@noop {} {\bibfield  {journal} {\bibinfo  {journal}
  {Science}\ }\textbf {\bibinfo {volume} {251}},\ \bibinfo {pages} {919}
  (\bibinfo {year} {1991})}\BibitemShut {NoStop}%
\bibitem [{\citenamefont {Gerischer}(1969)}]{gerischer1969charge}%
  \BibitemOpen
  \bibfield  {author} {\bibinfo {author} {\bibfnamefont {H.}~\bibnamefont
  {Gerischer}},\ }\href@noop {} {\bibfield  {journal} {\bibinfo  {journal}
  {Surf. Sci.}\ }\textbf {\bibinfo {volume} {18}},\ \bibinfo {pages} {97}
  (\bibinfo {year} {1969})}\BibitemShut {NoStop}%
\bibitem [{\citenamefont {Marcus}\ and\ \citenamefont
  {Sutin}(1986)}]{marcus1986relation}%
  \BibitemOpen
  \bibfield  {author} {\bibinfo {author} {\bibfnamefont {R.}~\bibnamefont
  {Marcus}}\ and\ \bibinfo {author} {\bibfnamefont {N.}~\bibnamefont {Sutin}},\
  }\href@noop {} {\bibfield  {journal} {\bibinfo  {journal} {Comments Inorg.
  Chem.}\ }\textbf {\bibinfo {volume} {5}},\ \bibinfo {pages} {119} (\bibinfo
  {year} {1986})}\BibitemShut {NoStop}%
\bibitem [{\citenamefont {Marcus}\ and\ \citenamefont
  {Sutin}(1975)}]{marcus1975electron}%
  \BibitemOpen
  \bibfield  {author} {\bibinfo {author} {\bibfnamefont {R.~A.}\ \bibnamefont
  {Marcus}}\ and\ \bibinfo {author} {\bibfnamefont {N.}~\bibnamefont {Sutin}},\
  }\href@noop {} {\bibfield  {journal} {\bibinfo  {journal} {Inorg. Chem.}\
  }\textbf {\bibinfo {volume} {14}},\ \bibinfo {pages} {213} (\bibinfo {year}
  {1975})}\BibitemShut {NoStop}%
\bibitem [{\citenamefont {Van~Duyne}\ and\ \citenamefont
  {Fischer}(1974)}]{van1974nonadiabatic}%
  \BibitemOpen
  \bibfield  {author} {\bibinfo {author} {\bibfnamefont {R.~P.}\ \bibnamefont
  {Van~Duyne}}\ and\ \bibinfo {author} {\bibfnamefont {S.~F.}\ \bibnamefont
  {Fischer}},\ }\href@noop {} {\bibfield  {journal} {\bibinfo  {journal} {Chem.
  Phys.}\ }\textbf {\bibinfo {volume} {5}},\ \bibinfo {pages} {183} (\bibinfo
  {year} {1974})}\BibitemShut {NoStop}%
\bibitem [{\citenamefont {Ulstrup}\ and\ \citenamefont
  {Jortner}(1975)}]{ulstrup1975effect}%
  \BibitemOpen
  \bibfield  {author} {\bibinfo {author} {\bibfnamefont {J.}~\bibnamefont
  {Ulstrup}}\ and\ \bibinfo {author} {\bibfnamefont {J.}~\bibnamefont
  {Jortner}},\ }\href@noop {} {\bibfield  {journal} {\bibinfo  {journal} {J.
  Chem. Phys.}\ }\textbf {\bibinfo {volume} {63}},\ \bibinfo {pages} {4358}
  (\bibinfo {year} {1975})}\BibitemShut {NoStop}%
\bibitem [{\citenamefont {Kestner}, \citenamefont {Logan},\ and\ \citenamefont
  {Jortner}(1974)}]{kestner1974thermal}%
  \BibitemOpen
  \bibfield  {author} {\bibinfo {author} {\bibfnamefont {N.~R.}\ \bibnamefont
  {Kestner}}, \bibinfo {author} {\bibfnamefont {J.}~\bibnamefont {Logan}}, \
  and\ \bibinfo {author} {\bibfnamefont {J.}~\bibnamefont {Jortner}},\
  }\href@noop {} {\bibfield  {journal} {\bibinfo  {journal} {J. Phys. Chem.}\
  }\textbf {\bibinfo {volume} {78}},\ \bibinfo {pages} {2148} (\bibinfo {year}
  {1974})}\BibitemShut {NoStop}%
\bibitem [{\citenamefont {Marcus}(1956)}]{marcus1956theory}%
  \BibitemOpen
  \bibfield  {author} {\bibinfo {author} {\bibfnamefont {R.~A.}\ \bibnamefont
  {Marcus}},\ }\href@noop {} {\bibfield  {journal} {\bibinfo  {journal} {J.
  Chem. Phys.}\ }\textbf {\bibinfo {volume} {24}},\ \bibinfo {pages} {966}
  (\bibinfo {year} {1956})}\BibitemShut {NoStop}%
\bibitem [{\citenamefont {Hopfield}(1974)}]{hopfield1974electron}%
  \BibitemOpen
  \bibfield  {author} {\bibinfo {author} {\bibfnamefont {J.~J.}\ \bibnamefont
  {Hopfield}},\ }\href@noop {} {\bibfield  {journal} {\bibinfo  {journal}
  {Proc. Natl. Acad. Sci. U.S.A.}\ }\textbf {\bibinfo {volume} {71}},\ \bibinfo
  {pages} {3640} (\bibinfo {year} {1974})}\BibitemShut {NoStop}%
\bibitem [{\citenamefont {Jortner}(1976)}]{jortner1976temperature}%
  \BibitemOpen
  \bibfield  {author} {\bibinfo {author} {\bibfnamefont {J.}~\bibnamefont
  {Jortner}},\ }\href@noop {} {\bibfield  {journal} {\bibinfo  {journal} {J.
  Chem. Phys.}\ }\textbf {\bibinfo {volume} {64}},\ \bibinfo {pages} {4860}
  (\bibinfo {year} {1976})}\BibitemShut {NoStop}%
\bibitem [{\citenamefont {Sumi}\ and\ \citenamefont
  {Marcus}(1986)}]{sumi1986dynamical}%
  \BibitemOpen
  \bibfield  {author} {\bibinfo {author} {\bibfnamefont {H.}~\bibnamefont
  {Sumi}}\ and\ \bibinfo {author} {\bibfnamefont {R.}~\bibnamefont {Marcus}},\
  }\href@noop {} {\bibfield  {journal} {\bibinfo  {journal} {J. Chem. Phys.}\
  }\textbf {\bibinfo {volume} {84}},\ \bibinfo {pages} {4894} (\bibinfo {year}
  {1986})}\BibitemShut {NoStop}%
\bibitem [{\citenamefont {Dou}\ \emph {et~al.}(2018)\citenamefont {Dou},
  \citenamefont {Schinabeck}, \citenamefont {Thoss},\ and\ \citenamefont
  {Subotnik}}]{dou2018broadened}%
  \BibitemOpen
  \bibfield  {author} {\bibinfo {author} {\bibfnamefont {W.}~\bibnamefont
  {Dou}}, \bibinfo {author} {\bibfnamefont {C.}~\bibnamefont {Schinabeck}},
  \bibinfo {author} {\bibfnamefont {M.}~\bibnamefont {Thoss}}, \ and\ \bibinfo
  {author} {\bibfnamefont {J.~E.}\ \bibnamefont {Subotnik}},\ }\href@noop {}
  {\bibfield  {journal} {\bibinfo  {journal} {J. Chem. Phys.}\ }\textbf
  {\bibinfo {volume} {148}},\ \bibinfo {pages} {102317} (\bibinfo {year}
  {2018})}\BibitemShut {NoStop}%
\bibitem [{\citenamefont {Kirchberg}, \citenamefont {Thorwart},\ and\
  \citenamefont {Nitzan}(2020)}]{kirchberg2020charge}%
  \BibitemOpen
  \bibfield  {author} {\bibinfo {author} {\bibfnamefont {H.}~\bibnamefont
  {Kirchberg}}, \bibinfo {author} {\bibfnamefont {M.}~\bibnamefont {Thorwart}},
  \ and\ \bibinfo {author} {\bibfnamefont {A.}~\bibnamefont {Nitzan}},\
  }\href@noop {} {\bibfield  {journal} {\bibinfo  {journal} {J. Phys. Chem.
  Lett.}\ }\textbf {\bibinfo {volume} {11}},\ \bibinfo {pages} {1729} (\bibinfo
  {year} {2020})}\BibitemShut {NoStop}%
\bibitem [{\citenamefont {Landauer}(1957)}]{landauer1957spatial}%
  \BibitemOpen
  \bibfield  {author} {\bibinfo {author} {\bibfnamefont {R.}~\bibnamefont
  {Landauer}},\ }\href@noop {} {\bibfield  {journal} {\bibinfo  {journal} {IBM
  J. Res. Dev.}\ }\textbf {\bibinfo {volume} {1}},\ \bibinfo {pages} {223}
  (\bibinfo {year} {1957})}\BibitemShut {NoStop}%
\bibitem [{\citenamefont {Imry}\ and\ \citenamefont
  {Landauer}(1999)}]{imry1999conductance}%
  \BibitemOpen
  \bibfield  {author} {\bibinfo {author} {\bibfnamefont {Y.}~\bibnamefont
  {Imry}}\ and\ \bibinfo {author} {\bibfnamefont {R.}~\bibnamefont
  {Landauer}},\ }\href@noop {} {\bibfield  {journal} {\bibinfo  {journal} {Rev.
  Mod. Phys.}\ }\textbf {\bibinfo {volume} {71}},\ \bibinfo {pages} {S306}
  (\bibinfo {year} {1999})}\BibitemShut {NoStop}%
\bibitem [{\citenamefont {Zimbovskaya}(2013)}]{zimbovskaya2013transport}%
  \BibitemOpen
  \bibfield  {author} {\bibinfo {author} {\bibfnamefont {N.~A.}\ \bibnamefont
  {Zimbovskaya}},\ }\href@noop {} {\emph {\bibinfo {title} {Transport
  properties of molecular junctions}}},\ Vol.\ \bibinfo {volume} {254}\
  (\bibinfo  {publisher} {Springer},\ \bibinfo {year} {2013})\BibitemShut
  {NoStop}%
\bibitem [{\citenamefont {Nitzan}(2006)}]{nitzan2006chemical}%
  \BibitemOpen
  \bibfield  {author} {\bibinfo {author} {\bibfnamefont {A.}~\bibnamefont
  {Nitzan}},\ }\href@noop {} {\emph {\bibinfo {title} {Chemical dynamics in
  condensed phases: relaxation, transfer and reactions in condensed molecular
  systems}}}\ (\bibinfo  {publisher} {{Oxford University Press}},\ \bibinfo
  {year} {2006})\BibitemShut {NoStop}%
\bibitem [{\citenamefont {Esposito}\ and\ \citenamefont
  {Galperin}(2009)}]{esposito2009transport}%
  \BibitemOpen
  \bibfield  {author} {\bibinfo {author} {\bibfnamefont {M.}~\bibnamefont
  {Esposito}}\ and\ \bibinfo {author} {\bibfnamefont {M.}~\bibnamefont
  {Galperin}},\ }\href@noop {} {\bibfield  {journal} {\bibinfo  {journal}
  {Phys. Rev. B}\ }\textbf {\bibinfo {volume} {79}},\ \bibinfo {pages} {205303}
  (\bibinfo {year} {2009})}\BibitemShut {NoStop}%
\bibitem [{\citenamefont {Breit}\ and\ \citenamefont
  {Wigner}(1936)}]{breit1936capture}%
  \BibitemOpen
  \bibfield  {author} {\bibinfo {author} {\bibfnamefont {G.}~\bibnamefont
  {Breit}}\ and\ \bibinfo {author} {\bibfnamefont {E.}~\bibnamefont {Wigner}},\
  }\href@noop {} {\bibfield  {journal} {\bibinfo  {journal} {Phys. Rev.}\
  }\textbf {\bibinfo {volume} {49}},\ \bibinfo {pages} {519} (\bibinfo {year}
  {1936})}\BibitemShut {NoStop}%
\bibitem [{\citenamefont {Landauer}(1989)}]{landauer1989conductance}%
  \BibitemOpen
  \bibfield  {author} {\bibinfo {author} {\bibfnamefont {R.}~\bibnamefont
  {Landauer}},\ }\href@noop {} {\bibfield  {journal} {\bibinfo  {journal} {J.
  Phys.: Condens. Matter}\ }\textbf {\bibinfo {volume} {1}},\ \bibinfo {pages}
  {8099} (\bibinfo {year} {1989})}\BibitemShut {NoStop}%
\bibitem [{\citenamefont {Gosavi}\ and\ \citenamefont
  {Marcus}(2000)}]{gosavi2000nonadiabatic}%
  \BibitemOpen
  \bibfield  {author} {\bibinfo {author} {\bibfnamefont {S.}~\bibnamefont
  {Gosavi}}\ and\ \bibinfo {author} {\bibfnamefont {R.~A.}\ \bibnamefont
  {Marcus}},\ }\href@noop {} {\bibfield  {journal} {\bibinfo  {journal} {J.
  Phys. Chem. B}\ }\textbf {\bibinfo {volume} {104}},\ \bibinfo {pages} {2067}
  (\bibinfo {year} {2000})}\BibitemShut {NoStop}%
\bibitem [{\citenamefont {Migliore}, \citenamefont {Schiff},\ and\
  \citenamefont {Nitzan}(2012)}]{migliore2012relationship}%
  \BibitemOpen
  \bibfield  {author} {\bibinfo {author} {\bibfnamefont {A.}~\bibnamefont
  {Migliore}}, \bibinfo {author} {\bibfnamefont {P.}~\bibnamefont {Schiff}}, \
  and\ \bibinfo {author} {\bibfnamefont {A.}~\bibnamefont {Nitzan}},\
  }\href@noop {} {\bibfield  {journal} {\bibinfo  {journal} {Phys. Chem. Chem.
  Phys.}\ }\textbf {\bibinfo {volume} {14}},\ \bibinfo {pages} {13746}
  (\bibinfo {year} {2012})}\BibitemShut {NoStop}%
\bibitem [{Note1()}]{Note1}%
  \BibitemOpen
  \bibinfo {note} {It is interesting to note that a Gaussian profile is
  sometimes introduced \protect \textit {ad hoc} into the Landauer framework in
  order to explain the experimentally-observed behavior.\cite
  {chen2017molecular}}\BibitemShut {NoStop}%
\bibitem [{\citenamefont {Armstrong}(1967)}]{armstrong1967spectrum}%
  \BibitemOpen
  \bibfield  {author} {\bibinfo {author} {\bibfnamefont {B.~H.}\ \bibnamefont
  {Armstrong}},\ }\href@noop {} {\bibfield  {journal} {\bibinfo  {journal} {J.
  Quant. Spectrosc. Radiat. Transf.}\ }\textbf {\bibinfo {volume} {7}},\
  \bibinfo {pages} {61} (\bibinfo {year} {1967})}\BibitemShut {NoStop}%
\bibitem [{\citenamefont {Beratan}, \citenamefont {Naaman},\ and\ \citenamefont
  {Waldeck}(2017)}]{beratan2017charge}%
  \BibitemOpen
  \bibfield  {author} {\bibinfo {author} {\bibfnamefont {D.~N.}\ \bibnamefont
  {Beratan}}, \bibinfo {author} {\bibfnamefont {R.}~\bibnamefont {Naaman}}, \
  and\ \bibinfo {author} {\bibfnamefont {D.~H.}\ \bibnamefont {Waldeck}},\
  }\href@noop {} {\bibfield  {journal} {\bibinfo  {journal} {Current Opinion in
  Electrochemistry}\ }\textbf {\bibinfo {volume} {4}},\ \bibinfo {pages} {175}
  (\bibinfo {year} {2017})}\BibitemShut {NoStop}%
\bibitem [{\citenamefont {Kim}, \citenamefont {Kilgour},\ and\ \citenamefont
  {Segal}(2016)}]{kim2016intermediate}%
  \BibitemOpen
  \bibfield  {author} {\bibinfo {author} {\bibfnamefont {H.}~\bibnamefont
  {Kim}}, \bibinfo {author} {\bibfnamefont {M.}~\bibnamefont {Kilgour}}, \ and\
  \bibinfo {author} {\bibfnamefont {D.}~\bibnamefont {Segal}},\ }\href@noop {}
  {\bibfield  {journal} {\bibinfo  {journal} {The Journal of Physical Chemistry
  C}\ }\textbf {\bibinfo {volume} {120}},\ \bibinfo {pages} {23951} (\bibinfo
  {year} {2016})}\BibitemShut {NoStop}%
\bibitem [{\citenamefont {Wierzbinski}\ \emph {et~al.}(2013)\citenamefont
  {Wierzbinski}, \citenamefont {Venkatramani}, \citenamefont {Davis},
  \citenamefont {Bezer}, \citenamefont {Kong}, \citenamefont {Xing},
  \citenamefont {Borguet}, \citenamefont {Achim}, \citenamefont {Beratan},\
  and\ \citenamefont {Waldeck}}]{wierzbinski2013single}%
  \BibitemOpen
  \bibfield  {author} {\bibinfo {author} {\bibfnamefont {E.}~\bibnamefont
  {Wierzbinski}}, \bibinfo {author} {\bibfnamefont {R.}~\bibnamefont
  {Venkatramani}}, \bibinfo {author} {\bibfnamefont {K.~L.}\ \bibnamefont
  {Davis}}, \bibinfo {author} {\bibfnamefont {S.}~\bibnamefont {Bezer}},
  \bibinfo {author} {\bibfnamefont {J.}~\bibnamefont {Kong}}, \bibinfo {author}
  {\bibfnamefont {Y.}~\bibnamefont {Xing}}, \bibinfo {author} {\bibfnamefont
  {E.}~\bibnamefont {Borguet}}, \bibinfo {author} {\bibfnamefont
  {C.}~\bibnamefont {Achim}}, \bibinfo {author} {\bibfnamefont {D.~N.}\
  \bibnamefont {Beratan}}, \ and\ \bibinfo {author} {\bibfnamefont {D.~H.}\
  \bibnamefont {Waldeck}},\ }\href@noop {} {\bibfield  {journal} {\bibinfo
  {journal} {ACS Nano}\ }\textbf {\bibinfo {volume} {7}},\ \bibinfo {pages}
  {5391} (\bibinfo {year} {2013})}\BibitemShut {NoStop}%
\bibitem [{\citenamefont {Dauphin-Ducharme}, \citenamefont
  {Arroyo-Curr{\'a}s},\ and\ \citenamefont {Plaxco}(2019)}]{dauphin2019high}%
  \BibitemOpen
  \bibfield  {author} {\bibinfo {author} {\bibfnamefont {P.}~\bibnamefont
  {Dauphin-Ducharme}}, \bibinfo {author} {\bibfnamefont {N.}~\bibnamefont
  {Arroyo-Curr{\'a}s}}, \ and\ \bibinfo {author} {\bibfnamefont {K.~W.}\
  \bibnamefont {Plaxco}},\ }\href@noop {} {\bibfield  {journal} {\bibinfo
  {journal} {J. Am. Chem. Soc.}\ }\textbf {\bibinfo {volume} {141}},\ \bibinfo
  {pages} {1304} (\bibinfo {year} {2019})}\BibitemShut {NoStop}%
\bibitem [{\citenamefont {Sowa}, \citenamefont {Mol},\ and\ \citenamefont
  {Gauger}(2019)}]{sowa2019marcus}%
  \BibitemOpen
  \bibfield  {author} {\bibinfo {author} {\bibfnamefont {J.~K.}\ \bibnamefont
  {Sowa}}, \bibinfo {author} {\bibfnamefont {J.~A.}\ \bibnamefont {Mol}}, \
  and\ \bibinfo {author} {\bibfnamefont {E.~M.}\ \bibnamefont {Gauger}},\
  }\href@noop {} {\bibfield  {journal} {\bibinfo  {journal} {J. Phys. Chem. C}\
  }\textbf {\bibinfo {volume} {123}},\ \bibinfo {pages} {4103} (\bibinfo {year}
  {2019})}\BibitemShut {NoStop}%
\bibitem [{Note2()}]{Note2}%
  \BibitemOpen
  \bibinfo {note} {Naturally, this does not hold for the approximation of
  $K_\pm (\epsilon )$ given in Eq.~\protect \textup {\hbox {\mathsurround \z@
  \protect \normalfont (\ignorespaces \ref {approxmm}\unskip \@@italiccorr )}}
  which is valid only on the part of the energy domain.}\BibitemShut {Stop}%
\bibitem [{\citenamefont {Choi}, \citenamefont {Kim},\ and\ \citenamefont
  {Frisbie}(2008)}]{choi2008electrical}%
  \BibitemOpen
  \bibfield  {author} {\bibinfo {author} {\bibfnamefont {S.~H.}\ \bibnamefont
  {Choi}}, \bibinfo {author} {\bibfnamefont {B.}~\bibnamefont {Kim}}, \ and\
  \bibinfo {author} {\bibfnamefont {C.~D.}\ \bibnamefont {Frisbie}},\
  }\href@noop {} {\bibfield  {journal} {\bibinfo  {journal} {Science}\ }\textbf
  {\bibinfo {volume} {320}},\ \bibinfo {pages} {1482} (\bibinfo {year}
  {2008})}\BibitemShut {NoStop}%
\bibitem [{\citenamefont {Beebe}\ \emph {et~al.}(2006)\citenamefont {Beebe},
  \citenamefont {Kim}, \citenamefont {Gadzuk}, \citenamefont {Frisbie},\ and\
  \citenamefont {Kushmerick}}]{beebe2006transition}%
  \BibitemOpen
  \bibfield  {author} {\bibinfo {author} {\bibfnamefont {J.~M.}\ \bibnamefont
  {Beebe}}, \bibinfo {author} {\bibfnamefont {B.}~\bibnamefont {Kim}}, \bibinfo
  {author} {\bibfnamefont {J.~W.}\ \bibnamefont {Gadzuk}}, \bibinfo {author}
  {\bibfnamefont {C.~D.}\ \bibnamefont {Frisbie}}, \ and\ \bibinfo {author}
  {\bibfnamefont {J.~G.}\ \bibnamefont {Kushmerick}},\ }\href@noop {}
  {\bibfield  {journal} {\bibinfo  {journal} {Phys. Rev. Lett.}\ }\textbf
  {\bibinfo {volume} {97}},\ \bibinfo {pages} {026801} (\bibinfo {year}
  {2006})}\BibitemShut {NoStop}%
\bibitem [{\citenamefont {Wang}, \citenamefont {Lee},\ and\ \citenamefont
  {Reed}(2003)}]{wang2003mechanism}%
  \BibitemOpen
  \bibfield  {author} {\bibinfo {author} {\bibfnamefont {W.}~\bibnamefont
  {Wang}}, \bibinfo {author} {\bibfnamefont {T.}~\bibnamefont {Lee}}, \ and\
  \bibinfo {author} {\bibfnamefont {M.~A.}\ \bibnamefont {Reed}},\ }\href@noop
  {} {\bibfield  {journal} {\bibinfo  {journal} {Phys. Rev. B}\ }\textbf
  {\bibinfo {volume} {68}},\ \bibinfo {pages} {035416} (\bibinfo {year}
  {2003})}\BibitemShut {NoStop}%
\bibitem [{\citenamefont {Wold}\ and\ \citenamefont
  {Frisbie}(2001)}]{wold2001fabrication}%
  \BibitemOpen
  \bibfield  {author} {\bibinfo {author} {\bibfnamefont {D.~J.}\ \bibnamefont
  {Wold}}\ and\ \bibinfo {author} {\bibfnamefont {C.~D.}\ \bibnamefont
  {Frisbie}},\ }\href@noop {} {\bibfield  {journal} {\bibinfo  {journal} {J.
  Am. Chem. Soc.}\ }\textbf {\bibinfo {volume} {123}},\ \bibinfo {pages} {5549}
  (\bibinfo {year} {2001})}\BibitemShut {NoStop}%
\bibitem [{\citenamefont {Simmons}(1963)}]{simmons1963generalized}%
  \BibitemOpen
  \bibfield  {author} {\bibinfo {author} {\bibfnamefont {J.~G.}\ \bibnamefont
  {Simmons}},\ }\href@noop {} {\bibfield  {journal} {\bibinfo  {journal} {J.
  Appl. Phys.}\ }\textbf {\bibinfo {volume} {34}},\ \bibinfo {pages} {1793}
  (\bibinfo {year} {1963})}\BibitemShut {NoStop}%
\bibitem [{\citenamefont {Frisenda}\ and\ \citenamefont {van~der
  Zant}(2016)}]{frisenda2016transition}%
  \BibitemOpen
  \bibfield  {author} {\bibinfo {author} {\bibfnamefont {R.}~\bibnamefont
  {Frisenda}}\ and\ \bibinfo {author} {\bibfnamefont {H.~S.~J.}\ \bibnamefont
  {van~der Zant}},\ }\href@noop {} {\bibfield  {journal} {\bibinfo  {journal}
  {Phys. Rev. Lett.}\ }\textbf {\bibinfo {volume} {117}},\ \bibinfo {pages}
  {126804} (\bibinfo {year} {2016})}\BibitemShut {NoStop}%
\bibitem [{\citenamefont {Capozzi}\ \emph {et~al.}(2015)\citenamefont
  {Capozzi}, \citenamefont {Xia}, \citenamefont {Adak}, \citenamefont {Dell},
  \citenamefont {Liu}, \citenamefont {Taylor}, \citenamefont {Neaton},
  \citenamefont {Campos},\ and\ \citenamefont
  {Venkataraman}}]{capozzi2015single}%
  \BibitemOpen
  \bibfield  {author} {\bibinfo {author} {\bibfnamefont {B.}~\bibnamefont
  {Capozzi}}, \bibinfo {author} {\bibfnamefont {J.}~\bibnamefont {Xia}},
  \bibinfo {author} {\bibfnamefont {O.}~\bibnamefont {Adak}}, \bibinfo {author}
  {\bibfnamefont {E.~J.}\ \bibnamefont {Dell}}, \bibinfo {author}
  {\bibfnamefont {Z.-F.}\ \bibnamefont {Liu}}, \bibinfo {author} {\bibfnamefont
  {J.~C.}\ \bibnamefont {Taylor}}, \bibinfo {author} {\bibfnamefont {J.~B.}\
  \bibnamefont {Neaton}}, \bibinfo {author} {\bibfnamefont {L.~M.}\
  \bibnamefont {Campos}}, \ and\ \bibinfo {author} {\bibfnamefont
  {L.}~\bibnamefont {Venkataraman}},\ }\href@noop {} {\bibfield  {journal}
  {\bibinfo  {journal} {Nat. Nanotechnol.}\ }\textbf {\bibinfo {volume} {10}},\
  \bibinfo {pages} {522} (\bibinfo {year} {2015})}\BibitemShut {NoStop}%
\bibitem [{\citenamefont {Datta}\ \emph {et~al.}(1997)\citenamefont {Datta},
  \citenamefont {Tian}, \citenamefont {Hong}, \citenamefont {Reifenberger},
  \citenamefont {Henderson},\ and\ \citenamefont {Kubiak}}]{datta1997current}%
  \BibitemOpen
  \bibfield  {author} {\bibinfo {author} {\bibfnamefont {S.}~\bibnamefont
  {Datta}}, \bibinfo {author} {\bibfnamefont {W.}~\bibnamefont {Tian}},
  \bibinfo {author} {\bibfnamefont {S.}~\bibnamefont {Hong}}, \bibinfo {author}
  {\bibfnamefont {R.}~\bibnamefont {Reifenberger}}, \bibinfo {author}
  {\bibfnamefont {J.~I.}\ \bibnamefont {Henderson}}, \ and\ \bibinfo {author}
  {\bibfnamefont {C.~P.}\ \bibnamefont {Kubiak}},\ }\href@noop {} {\bibfield
  {journal} {\bibinfo  {journal} {Phys. Rev. Lett.}\ }\textbf {\bibinfo
  {volume} {79}},\ \bibinfo {pages} {2530} (\bibinfo {year}
  {1997})}\BibitemShut {NoStop}%
\bibitem [{\citenamefont {Koch}\ and\ \citenamefont {von
  Oppen}(2005)}]{koch2005franck}%
  \BibitemOpen
  \bibfield  {author} {\bibinfo {author} {\bibfnamefont {J.}~\bibnamefont
  {Koch}}\ and\ \bibinfo {author} {\bibfnamefont {F.}~\bibnamefont {von
  Oppen}},\ }\href@noop {} {\bibfield  {journal} {\bibinfo  {journal} {Phys.
  Rev. Lett.}\ }\textbf {\bibinfo {volume} {94}},\ \bibinfo {pages} {206804}
  (\bibinfo {year} {2005})}\BibitemShut {NoStop}%
\bibitem [{\citenamefont {Bevan}\ \emph {et~al.}(2018)\citenamefont {Bevan},
  \citenamefont {Roy-Gobeil}, \citenamefont {Miyahara},\ and\ \citenamefont
  {Grutter}}]{bevan2018relating}%
  \BibitemOpen
  \bibfield  {author} {\bibinfo {author} {\bibfnamefont {K.~H.}\ \bibnamefont
  {Bevan}}, \bibinfo {author} {\bibfnamefont {A.}~\bibnamefont {Roy-Gobeil}},
  \bibinfo {author} {\bibfnamefont {Y.}~\bibnamefont {Miyahara}}, \ and\
  \bibinfo {author} {\bibfnamefont {P.}~\bibnamefont {Grutter}},\ }\href@noop
  {} {\bibfield  {journal} {\bibinfo  {journal} {J. Chem. Phys.}\ }\textbf
  {\bibinfo {volume} {149}},\ \bibinfo {pages} {104109} (\bibinfo {year}
  {2018})}\BibitemShut {NoStop}%
\bibitem [{\citenamefont {Marrosu}\ \emph {et~al.}(1990)\citenamefont
  {Marrosu}, \citenamefont {Rodante}, \citenamefont {Trazza},\ and\
  \citenamefont {Greci}}]{marrosu1990reaction}%
  \BibitemOpen
  \bibfield  {author} {\bibinfo {author} {\bibfnamefont {G.}~\bibnamefont
  {Marrosu}}, \bibinfo {author} {\bibfnamefont {F.}~\bibnamefont {Rodante}},
  \bibinfo {author} {\bibfnamefont {A.}~\bibnamefont {Trazza}}, \ and\ \bibinfo
  {author} {\bibfnamefont {L.}~\bibnamefont {Greci}},\ }\href@noop {}
  {\bibfield  {journal} {\bibinfo  {journal} {Thermochimica acta}\ }\textbf
  {\bibinfo {volume} {168}},\ \bibinfo {pages} {59} (\bibinfo {year}
  {1990})}\BibitemShut {NoStop}%
\bibitem [{\citenamefont {Komaguchi}\ \emph {et~al.}(1991)\citenamefont
  {Komaguchi}, \citenamefont {Hatsusegawa}, \citenamefont {Kitani},\ and\
  \citenamefont {Sasaki}}]{komaguchi1991entropy}%
  \BibitemOpen
  \bibfield  {author} {\bibinfo {author} {\bibfnamefont {K.}~\bibnamefont
  {Komaguchi}}, \bibinfo {author} {\bibfnamefont {Y.}~\bibnamefont
  {Hatsusegawa}}, \bibinfo {author} {\bibfnamefont {A.}~\bibnamefont {Kitani}},
  \ and\ \bibinfo {author} {\bibfnamefont {K.}~\bibnamefont {Sasaki}},\
  }\href@noop {} {\bibfield  {journal} {\bibinfo  {journal} {Bull. Chem. Soc.
  Jpn.}\ }\textbf {\bibinfo {volume} {64}},\ \bibinfo {pages} {2686} (\bibinfo
  {year} {1991})}\BibitemShut {NoStop}%
\bibitem [{\citenamefont {Svaan}\ and\ \citenamefont
  {Parker}(1984)}]{svaan1984temperature}%
  \BibitemOpen
  \bibfield  {author} {\bibinfo {author} {\bibfnamefont {M.}~\bibnamefont
  {Svaan}}\ and\ \bibinfo {author} {\bibfnamefont {V.~D.}\ \bibnamefont
  {Parker}},\ }\href@noop {} {\bibfield  {journal} {\bibinfo  {journal} {Acta
  Chem. Scand. B}\ }\textbf {\bibinfo {volume} {38}},\ \bibinfo {pages} {759}
  (\bibinfo {year} {1984})}\BibitemShut {NoStop}%
\bibitem [{\citenamefont {Miller}, \citenamefont {Beitz},\ and\ \citenamefont
  {Huddleston}(1984)}]{miller1984effect}%
  \BibitemOpen
  \bibfield  {author} {\bibinfo {author} {\bibfnamefont {J.~R.}\ \bibnamefont
  {Miller}}, \bibinfo {author} {\bibfnamefont {J.~V.}\ \bibnamefont {Beitz}}, \
  and\ \bibinfo {author} {\bibfnamefont {R.~K.}\ \bibnamefont {Huddleston}},\
  }\href@noop {} {\bibfield  {journal} {\bibinfo  {journal} {J. Am. Chem.
  Soc.}\ }\textbf {\bibinfo {volume} {106}},\ \bibinfo {pages} {5057} (\bibinfo
  {year} {1984})}\BibitemShut {NoStop}%
\bibitem [{\citenamefont {Closs}\ \emph {et~al.}(1986)\citenamefont {Closs},
  \citenamefont {Calcaterra}, \citenamefont {Green}, \citenamefont {Penfield},\
  and\ \citenamefont {Miller}}]{closs1986distance}%
  \BibitemOpen
  \bibfield  {author} {\bibinfo {author} {\bibfnamefont {G.~L.}\ \bibnamefont
  {Closs}}, \bibinfo {author} {\bibfnamefont {L.~T.}\ \bibnamefont
  {Calcaterra}}, \bibinfo {author} {\bibfnamefont {N.~J.}\ \bibnamefont
  {Green}}, \bibinfo {author} {\bibfnamefont {K.~W.}\ \bibnamefont {Penfield}},
  \ and\ \bibinfo {author} {\bibfnamefont {J.~R.}\ \bibnamefont {Miller}},\
  }\href@noop {} {\bibfield  {journal} {\bibinfo  {journal} {J. Phys. Chem.}\
  }\textbf {\bibinfo {volume} {90}},\ \bibinfo {pages} {3673} (\bibinfo {year}
  {1986})}\BibitemShut {NoStop}%
\bibitem [{Note3()}]{Note3}%
  \BibitemOpen
  \bibinfo {note} {However, an \protect \textit {ad hoc} replacement of the
  molecule-lead hopping rates (in the conventional rate equation or quantum
  master equation approaches) with the expressions developed here will yield a
  theory that does not recover the exact Landauer result for molecular systems
  with more than one site.}\BibitemShut {Stop}%
\bibitem [{\citenamefont {Abrarov}\ and\ \citenamefont
  {Quine}(2015)}]{abrarov2015accurate}%
  \BibitemOpen
  \bibfield  {author} {\bibinfo {author} {\bibfnamefont {S.~M.}\ \bibnamefont
  {Abrarov}}\ and\ \bibinfo {author} {\bibfnamefont {B.~M.}\ \bibnamefont
  {Quine}},\ }\href@noop {} {\bibfield  {journal} {\bibinfo  {journal} {J.
  Math. Res.}\ }\textbf {\bibinfo {volume} {7}},\ \bibinfo {pages} {44}
  (\bibinfo {year} {2015})}\BibitemShut {NoStop}%
\end{thebibliography}
\end{document}